\documentclass[%
aps,pre,reprint,superscriptaddress
]{revtex4-2}
\usepackage{changes}
\usepackage[english]{babel}

\usepackage{amsmath}
\usepackage{amsfonts}
\usepackage{graphicx}
\usepackage[colorlinks=true, allcolors=blue]{hyperref}
\usepackage{float}
\usepackage{comment}
\usepackage{caption}

\usepackage{xcolor}

\usepackage{subcaption}
\newcommand{\boldx}{\boldsymbol{x}}
\newcommand{\boldxgamma}{\bar{\boldsymbol{x}}(s)} 
\newcommand{\xgamma}{ \bar x}
\newcommand{\ygamma}{ \bar y}
\newcommand{\eunit}{\mathbf{e}}
\newcommand{\vBC}{v_\textsc{\tiny BC}}
\newcommand{\vSP}{v_\textsc{\tiny SP}}
\newcommand{\nablauu}{\nabla_{\Dot{\boldsymbol x}}\Dot{\boldsymbol{x}}}
\newcommand{\boldxbar}{ \bar{\boldsymbol{x}} }
\newcommand{\Prob}{\mathbb{P}}
\newcommand{\vshift}{v_\textsc{\tiny shifted}^\|}
\begin{document}

\title{Stochastic fluctuations of diluted pedestrian dynamics along curved paths}
\author{Geert G. M. van der Vleuten}
\affiliation{Department of Applied Physics and Science Education, Eindhoven University of Technology,  5600 MB Eindhoven, The Netherlands}
\author{Federico Toschi}
\affiliation{Department of Applied Physics and Science Education, Eindhoven University of Technology,  5600 MB Eindhoven, The Netherlands}
\affiliation{Consiglio Nazionale della Ricerche-IAC, Rome, Italy}
\author{Wil H. A. Schilders}
\affiliation{Department of Mathematics and Computer Science, Eindhoven University of Technology,  5600 MB Eindhoven, The Netherlands}
\author{Alessandro Corbetta}
\email{a.corbetta@tue.nl}
\affiliation{Department of Applied Physics and Science Education, Eindhoven University of Technology,  5600 MB Eindhoven, The Netherlands}

\begin{abstract}
As we walk towards our destinations, our trajectories are constantly influenced by the presence of obstacles and infrastructural elements: even in absence of crowding our paths are often curved. 
Over the last two decades pedestrian dynamics have been extensively studied aiming at quantitative models with both fundamental and technological relevance. Walking kinematics along straight paths have been experimentally investigated and quantitatively modeled in the diluted limit (i.e. in absence of  pedestrian-pedestrian interactions). 
It is natural to expect that models for straight paths may be an accurate approximations of the dynamics even for paths with curvature radii much larger than the size of a single person. Conversely, as paths curvature increase one may expect larger and larger deviations. As no clear experimental consensus has been reached yet in the literature, here we accurately and systematically investigate the effect of paths curvature on diluted pedestrian dynamics. Thanks to a extensive and highly accurate set of real-life measurements campaign, we derive and validate via a Langevin-like social-force model capable of quantitatively describing both averages and fluctuations. Leveraging on the differential geometric notion of covariant derivative, we generalize previous work by some of the authors, effectively casting a Langevin social-force model for the straight walking dynamics in a curved geometric setting. We deem this the necessary first step to understand  and model the more general and ubiquitous case of pedestrians following curved paths in the presence of crowd traffic.

\end{abstract}

\maketitle

\section{Introduction}

\begin{figure*}[ht]
    \centering
    \includegraphics[width=\textwidth]{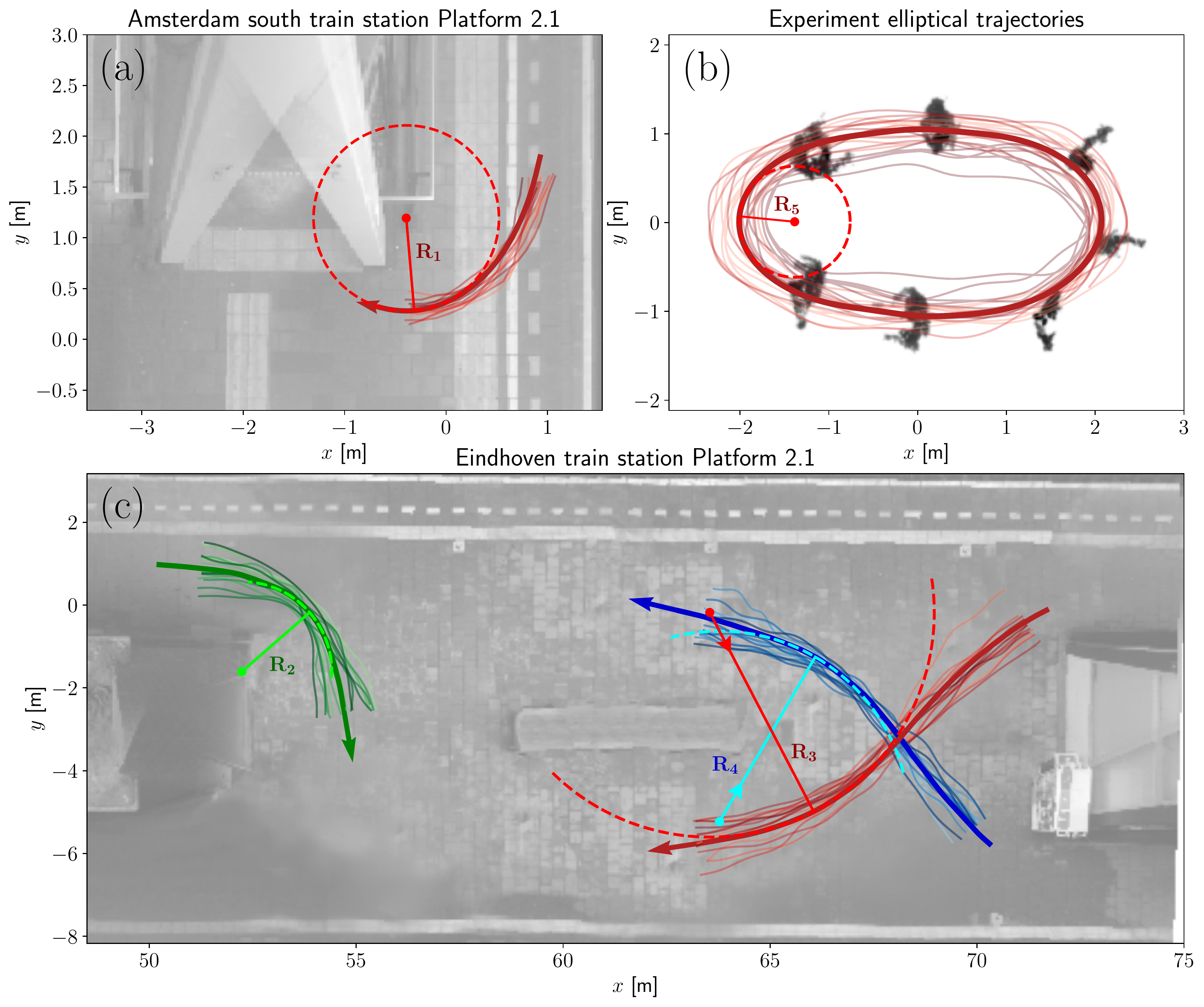}
    \caption{Few selected trajectories of pedestrians walking along different curved paths in different real-life locations. The three panels provide an overview of the measurement sites employed in this work, respectively in Amsterdam south train station (NL) (a), a laboratory experiment at Eindhoven University of Technology (NL) (where pedestrian were asked to walk along an elliptical path) (b), and at Eindhoven train station (NL) (c). The tangent circles at several points of the preferred-paths (average path across trajectory bundles, cf.~Appendix~\ref{Appendix: average}) are displayed with radii $R_1=0.92$m, $R_2=2.17$m, $R_3=5.42$m, $R_4=4.60$m and $R_5=0.63$, respectively.}
    \label{fig:trajectories trainstation}
\end{figure*}

As we walk towards our destinations, indoor or in open spaces, we typically prefer to follow the most direct (typically straight) path. 
Yet, obstacles, infrastructural elements, or crowd traffic~\cite{Parisi2016,Corbetta2018}, make our \textit{preferred paths} unavoidably \textit{curved} (cf. Fig.~\ref{fig:trajectories trainstation}). 
Additionally, trajectories invariably exhibit \textit{fluctuations} associated with sway and inter-subject variability.

 Over the last two decades, pedestrian kinematics has been extensively investigated experimentally~\cite{Corbetta2023,feliciani2021introduction}, and the motion of pedestrians walking along straight paths has been thoroughly analyzed and modeled (e.g.~\cite{zhang2014universal,zhang2012ordering, murakami2021mutual, Corbetta2018}). Especially in diluted conditions, i.e. in absence of pedestrian-pedestrian interactions, these analyses were capable of successfully modeling the dynamics, including the stochastic fluctuations around  average motions~\cite{Corbetta2017,corbetta2018path}. In the case of paths having curvature radii much larger than the scale of a single person, we expect models for straight dynamics to hold locally. In fact, under these conditions, paths can be reasonably well approximated as being locally straight. One may thus wonder under which conditions and how the known model for straight paths can be adapted to generic curved paths. %
Indeed, as paths curvatures increase, one may expect larger and larger deviations from the the assumption of a locally straight dynamics. No experimental consensus has yet been reached on how paths curvature affect pedestrians dynamics. 
Only few, and partially contradictory, studies are available on the topic. These report anti-correlation between velocity and curvature (with linear~\cite{Parisi2016} or power law trend~\cite{Hicheur2005}) or, even, an apparent absence of curvature effects~\cite{Ziemer2016}.

The aim of this work is to understand and to quantitatively model the dynamics of pedestrians walking along curved preferred paths, including averages and stochastic fluctuations, considering  a broad spectrum of curvature radii even as small as few pedestrian diameters.  We opt to address this outstanding issue restricting to crowd scenarios in the diluted limit. Thus, the environment is the only reason pedestrians opt for curved paths. We deem this setting the necessary first step towards the goal of understanding the generic case in which curved paths appear in combination with and as a consequence of  the overall crowd traffic. 

Understanding the kinematics of pedestrians is part of a challenging and broad multidisciplinary scientific effort with outstanding societal importance due to implications in crowd management~\cite{feliciani2021introduction} and urban design~\cite{Gerike2021}, and sharing deep fundamental challenges connected with active flowing matter and statistical physics~\cite{RevModPhys.85.1143,Corbetta2023}.
One of the main obstacles in fully understanding crowd flows is the inherent technical challenge of obtaining measurements with sufficient spatio-temporal accuracy and statistical resolution, fully capturing the large variability and complexity of pedestrian kinematics. 
Over the past years, experimental evidence on pedestrian behavior has been collected mostly in laboratory scenarios, allowing to probe average behavior, typically studied as a function of the pedestrian density (e.g.\,~\cite{DBLP:journals/ijon/BoltesS13}). 
Average behavior are usually encoded in so-called fundamental diagrams, connecting, e.g., pedestrian density with average velocity or fluxes~\cite{Vanumu2017}. Only more recently, accurate and privacy-respectful large-scale measurements in real-life conditions have become a possibility, either via custom setups developed in research environments~\cite{DBLP:journals/ijon/BoltesS13, seer2014kinects} and via commercial products~\cite{Pouw_Willems_2022}. Key have been three-dimensional computer vision approaches based on stereoscopic vision or LiDar-like approaches~\cite{brscic2013person,gabbana2022fluctuations}. Data acquisition with a 24/7 schedule in public locations has enabled the collection of highly-resolved, high statistics datasets (millions of trajectories), allowing statistical analyses up to rare events and opening new possibilities of model validation~\cite{Corbetta2023,willems2020pedestrian,zuriguel2014clogging,Brscic201477,PhysRevE.89.012811,pouw2020monitoring}.

In this paper, we use for the first time high-resolution tracking to collect wide trajectory datasets to investigate the diluted dynamics of pedestrians walking along curved paths.
We have performed large-scale data acquisition campaigns in Dutch train stations (Eindhoven, Amsterdam South) and laboratory experiments (in the Eindhoven University of Technology campus, NL). On these bases, we identify the effect of increasing curvature levels on   walking velocities, presenting a curvature-velocity fundamental diagram, which we enrich with measurements of the typical fluctuations.
This enables us to present  a Langevin-like model reproducing quantitatively the complete statistics of position and velocity as curvature changes. 
Our work generalizes the social force-like~\cite{Helbing1995} model presented in~\cite{Corbetta2017}, which quantitatively reproduces the diluted walking dynamics along straight paths. 
We effectively cast such a model to a curved geometry: even in absence of (social)-forces, pedestrians could follow curved trajectories. 
For this, we employ the language of differential geometry (in particular, through the notion of covariant derivative).  
On the basis of our data analysis, we extend the social-force terms to integrate curvature-dependent effects (with~radii down to $0.6$~m).

This paper is structured as follows: in Sect.~\ref{section:kinematics of curved walking paths in tubular neighborhoods} we introduce the geometric context of tubular neighborhoods of trajectories, central for the forthcoming analyses.  In Sect.~\ref{section:measurements}, we present the experimental data that we collected for our analyses, together with relevant technical references on data acquisition. Based on the measurements, in Sect.~\ref{section:curvature velocity fundamental diagram and fluctuations}, we present a curvature-velocity fundamental diagram, comparing a simple analytic model with measurements. In Sect.~\ref{section: Langevin-like model for curved tubular neighborhood}, we present our quantitative Langevin-like model, whose comparison with measurements is reported in the results Sect.~\ref{section:resulst}. 
A final discussion closes the paper. We opt to postpone most of the technical and formal details connected with differential geometry to the appendices.

\begin{figure*}[ht]
    \centering
    \begin{subfigure}{0.4\textwidth}
    \includegraphics[trim=3cm 4.5cm 2.2cm 1.1cm,clip,height=5.3cm]{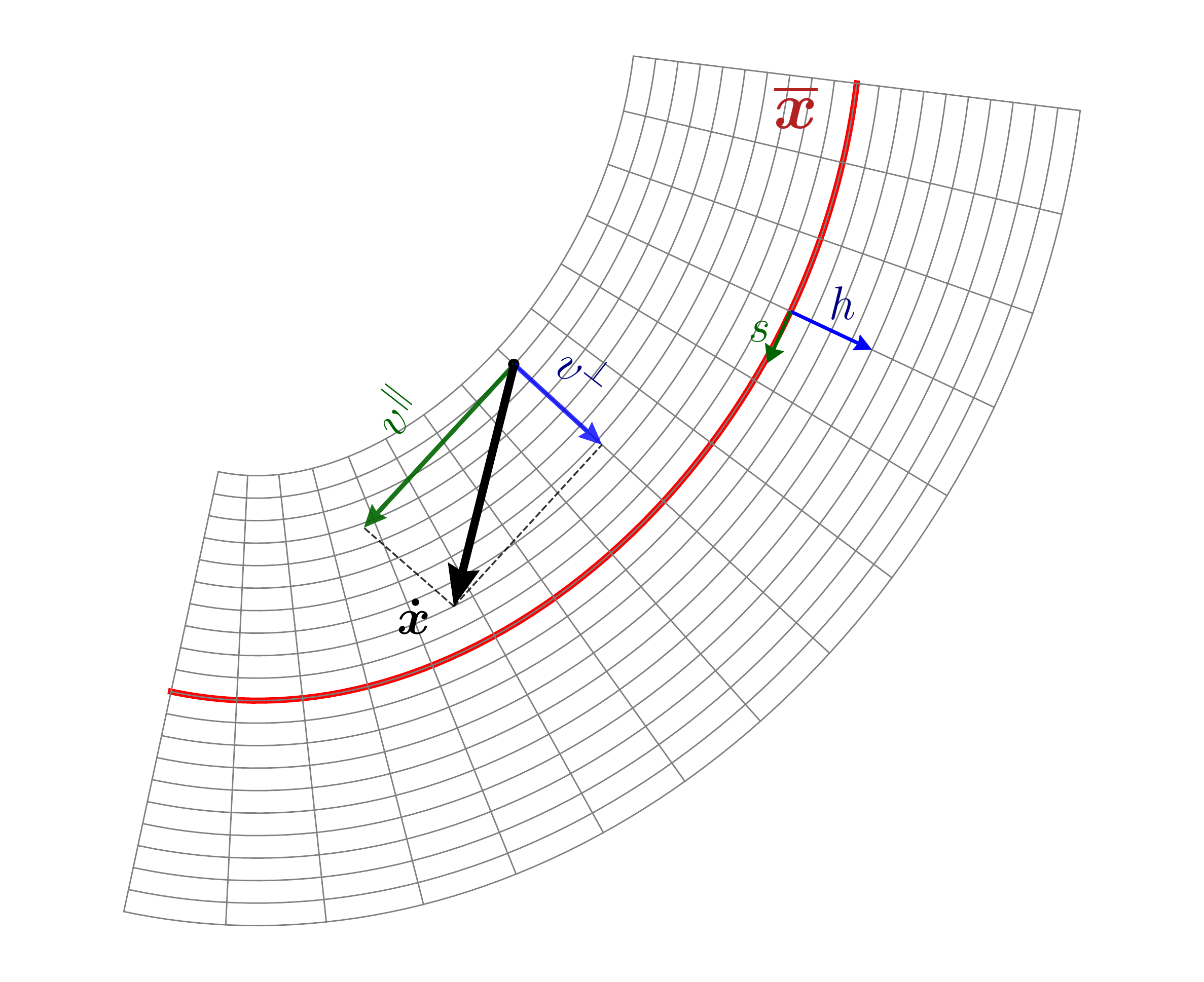}
    \caption{}\label{fig: sketch tubular neighborhood}
    \end{subfigure}
    \begin{subfigure}{0.55\textwidth}
    \includegraphics[trim=0cm 4.1cm 0cm 1cm,clip,height=5.3cm]{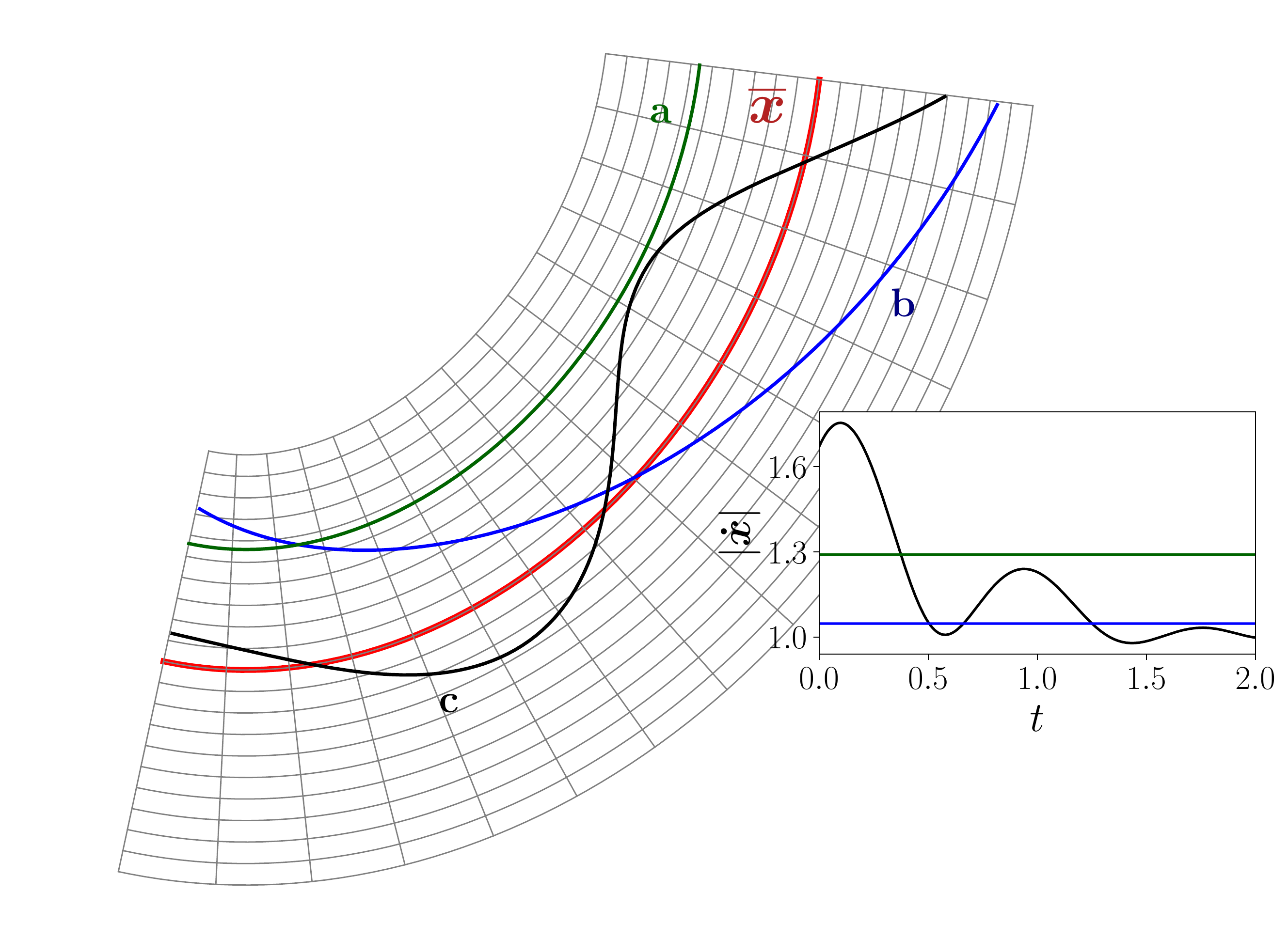}
    \caption{}\label{fig: sketch geodesic flow}
    \end{subfigure}
    \caption{(a) The average path of the Amsterdam train station measurements, curve $\boldxgamma$, indicated in red. The tubular neighborhood of $\boldxgamma$ is parameterized by a coordinate frame consisting of coordinate lines equidistant to $\boldxgamma$, defining the longitudinal direction, and coordinate lines perpendicular to $\boldxgamma$, defining the transversal direction. Coordinates $s$ and $h$ represent the coordinates in the longitudinal and transversal direction, respectively. As indicated, a velocity vector $V$ can be decomposed in a longitudinal and transversal component ($v^\|$ and $v^\perp$). (b) Trajectories $\mathbf{a}$, in green, and $\mathbf{b}$, in blue, are examples of geodesics (solutions of Eq.~\eqref{eq:geodesic flow}). Geodesics with initial velocity parallel to $\boldxgamma$, such as trajectory $\mathbf{a}$, remain parallel to $\boldxgamma$. The velocity magnitudes of the trajectories are plotted in the inset figure. All geodesics, such as trajectory $\mathbf{a}$ and $\mathbf{b}$, conserve kinetic energy. Trajectory $\mathbf{c}$, in black, is a solution of the geodesic equation disturbed by modeling forces (Eq.~\eqref{eq:main} without noise). Therefore, the trajectory is forced to oscillate around the path in a damped way. Furthermore, its longitudinal velocity converges to a desired value.}\label{fig: sketches}
\end{figure*}

\section{Kinematics of curved walking paths in tubular neighborhoods}\label{section:kinematics of curved walking paths in tubular neighborhoods}
We focus on bundles (i.e.~sets containing similarly shaped trajectories) of real-life pedestrian trajectories on the plane $\boldsymbol x=(x,y)$:
\begin{equation}\label{eq:bundle}
 \{t\mapsto \boldsymbol{x}_\nu(t) = x_\nu(t)\eunit_x + y_\nu(t)\eunit_y,\quad \nu=1,2,\ldots\},   
 \end{equation}
 where $\nu=1,2,\ldots$ serves as a trajectory index, $x_\nu(t), y_\nu(t)$ are the horizontal and vertical components of trajectory $\nu$ at time $t$, and $(\eunit_x,\eunit_y)$ is the (fixed) orthonormal base associated with the $(x,y)$ coordinates (cf. examples in Fig.~\ref{fig: sketches}). These trajectories connect predefined origin and destination, which are  separated by, e.g., obstacles or architectural fixtures. The need of bypassing these elements makes typical trajectories, and thus the whole bundle,  non-rectilinear. Due to sway and inter-subject variability, trajectories exhibit fluctuations. We analyze such fluctuations in reference with the average path of the bundle, 
 \begin{equation*}
\bar\boldx  = \boldxgamma = \xgamma(s)\eunit_x+ \ygamma(s)\eunit_y,   
 \end{equation*}
where the variable $s$ denotes a smooth monotonic parametrization. 
 We identify $\boldxgamma$ with the individual \textit{preferred} path, i.e.~the trajectory that each pedestrian aims at following. Examples of such average paths are reported as thick lines in Fig.~\ref{fig:trajectories trainstation}.
 We postpone the technicalities of the formal definition of the average path, $\boldxgamma$ (Eq.~\eqref{eq:bundle}), as a function of the trajectory bundle to Appendix~\ref{Appendix: average}.

We study fluctuations around $\boldxgamma$ considering its neighborhood. We employ coordinate lines parallel and normal to $\boldxgamma$ (Fig.~\ref{fig: sketch tubular neighborhood}), parametrized by the variables $s$ and $h$, respectively. As mentioned, $s$ increases as we move along $\boldxgamma$, whereas $h$ increases as we move in the orthogonal direction  (towards the local curvature center). We name $(\eunit_\|,\,\eunit_\perp)$ the local orthonormal base parallel to these directions. Note that curves defined by $h=const$ wrap around $\boldxgamma$ while remaining, in a sense, parallel to it. As such the $(s,h)$ parametrization of the $\boldxgamma$ neighborhood is usually named \textit{tubular}. For smooth $\boldxgamma$ and limited $h$, $(s,h)$ uniquely parameterize the tubular neighborhood (e.g.~\cite{gray2003tubes}). 
We unambiguously decompose velocities, $\dot\boldx = \dot x \eunit_x + \dot y \eunit_y$, applied at a point $\boldx$ in the neighborhood of $\boldxgamma$,  in a transversal,  $v^\perp$, and a longitudinal component, $v^\|$, respectively perpendicular and parallel to a local coordinate line ($h=const$). In formulas
\begin{equation}
    \dot \boldx = v^\|\eunit_\| +v^\perp\eunit_\perp.
\end{equation}
Further details on the parametrization of the tubular neighborhood are given in Appendix~\ref{appendix:diffgeo}.

Our analysis targets kinematic implications on pedestrian trajectories of the curvature of the preferred path. We consider the local curvature of $\boldxgamma$, $k(s)$.  
By definition, $k(s)$ is the reciprocal of the radius of the circle osculating $\boldxgamma$ and reads (e.g.~\cite{Courant1999})
\begin{equation}\label{curvature}
    k(s) = \frac{\xgamma^{\prime}(s) \ygamma^{\prime \prime}(s)-\xgamma^{\prime \prime}(s) \ygamma^{\prime}(s)}{\left[\left(\xgamma^{\prime}(s)\right)^{2}+\left(\ygamma^{\prime}(s)\right)^{2}\right]^{3 / 2}},
\end{equation}
where $\xgamma^{\prime}$ denotes the first derivative of the $x$ component of $\boldxgamma$ with respect to $s$ (the second derivative and operations on the $y$ component are written accordingly).

\section{Measurements}\label{section:measurements}
Our study leverages on trajectory datasets acquired via three large-scale pedestrian tracking campaigns all performed in The Netherlands. Our campaigns specifically took place in Amsterdam south train station (AMS), Eindhoven train station (EHV) and on the university campus in Eindhoven (TUE). All our data has been acquired in naturalistic condition (with the exception of the TUE campaign in which pedestrians have been instructed to roughly follow a given path) and in a fully privacy respectful manner. Commercial or research-grade overhead tracking sensors have been employed. Since we are interested in the dynamics of undisturbed pedestrians, we consider trajectories in low density conditions (i.e.~in absence of other neighboring pedestrians). 

In the following we provide a brief description of the datasets (for technical details about the average paths and the selection procedures, see Appendices~\ref{Appendix: average}-\ref{Appendix: data selection}).

\vspace{.2cm}
\noindent \textbf{Amsterdam south train station (AMS).} 
At this measurement location on platform 2.1, we consider high-resolution data in the vicinity of the staircase (Fig.~\ref{fig:trajectories trainstation}(a)) for the period spanning April 2020 to December 2020 (196 days).
Pedestrians arriving by train normally leave the platform via the staircase depicted in the middle. Thus we select some of the many trajectories of pedestrians turning from the platform towards the staircase. The strict selection criteria (Appendix~\ref{Appendix: data selection}) result in a selection of 2,700 measured trajectories in Amsterdam south train station.

The average path has a gradually increasing curvature and consequently a broad curvature spectrum with a radius of curvature ranging from 5 to 0.9 meters. The length of the average path is approximately 2 meters.

\vspace{.2cm}
\noindent \textbf{Eindhoven train station (EHV).} At the measurement domain within Eindhoven train station platform 2.1, measurements have been performed between April 2021 and September 2021 with a sample frequency of 10~Hz. We have chosen five winding paths in this train station as preferred paths as these are walked by many pedestrians. Additionally, all paths span wide curvature ranges. A top view of the platform with three preferred paths is shown in Fig.~\ref{fig:trajectories trainstation}(c). 

Totally 2,700 measured trajectories are selected in Eindhoven train station. The average paths in Eindhoven train station have lengths ranging from 4 to 10 meters. The minimum radius of curvature reached by the preferred paths in this station is 2.1 meters.

\vspace{.2cm}
\noindent \textbf{Eindhoven University of Technology (TUE).} This measurement campaign is conducted as an experiment at a large public area within the University campus in Eindhoven, the Netherlands in February 2019. During one minute, seven participants were asked to walk around two traffic cones, 3 meters apart, resulting in elliptical-like trajectories (Fig.~\ref{fig:trajectories trainstation}(b)). The pedestrians kept their distance to create diluted conditions. The average path has a broad curvature spectrum with a minimal radius of curvature around 0.6 meters. The measured trajectories are sampled with a frequency of 30~Hz (further technical information on this experimental setup based on overhead depth sensors are in~\cite{Pouw_Willems_2022}).

\section{Curvature-velocity fundamental diagram and fluctuations}\label{section:curvature velocity fundamental diagram and fluctuations}
We report here on the effect of the preferred path curvature on the average velocity in the diluted flow limit. We compare a closed-form theoretical model with high statistics measurements. These enable us to derive a fundamental diagram-like relation for average velocity and path curvature. 

Consistently with previous research~\cite{Parisi2016,Hicheur2005}, we observe that the walking velocity decreases with the curvature of the path. 

\begin{figure*}[ht]
    \begin{subfigure}{0.19\textwidth}    \includegraphics[trim=8.5cm 13cm 8.5cm 12.5cm,clip,width=\textwidth]{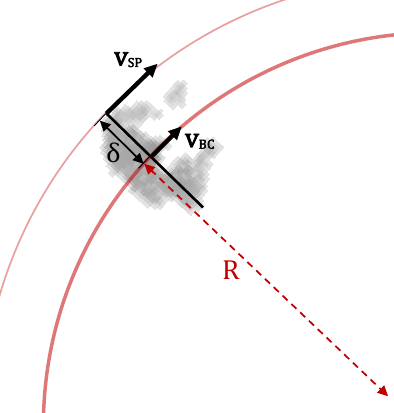}
    \caption{}\label{fig: sketch}
    \end{subfigure}
    \hfill
    \begin{subfigure}{0.8\textwidth}    \includegraphics[width=\textwidth]{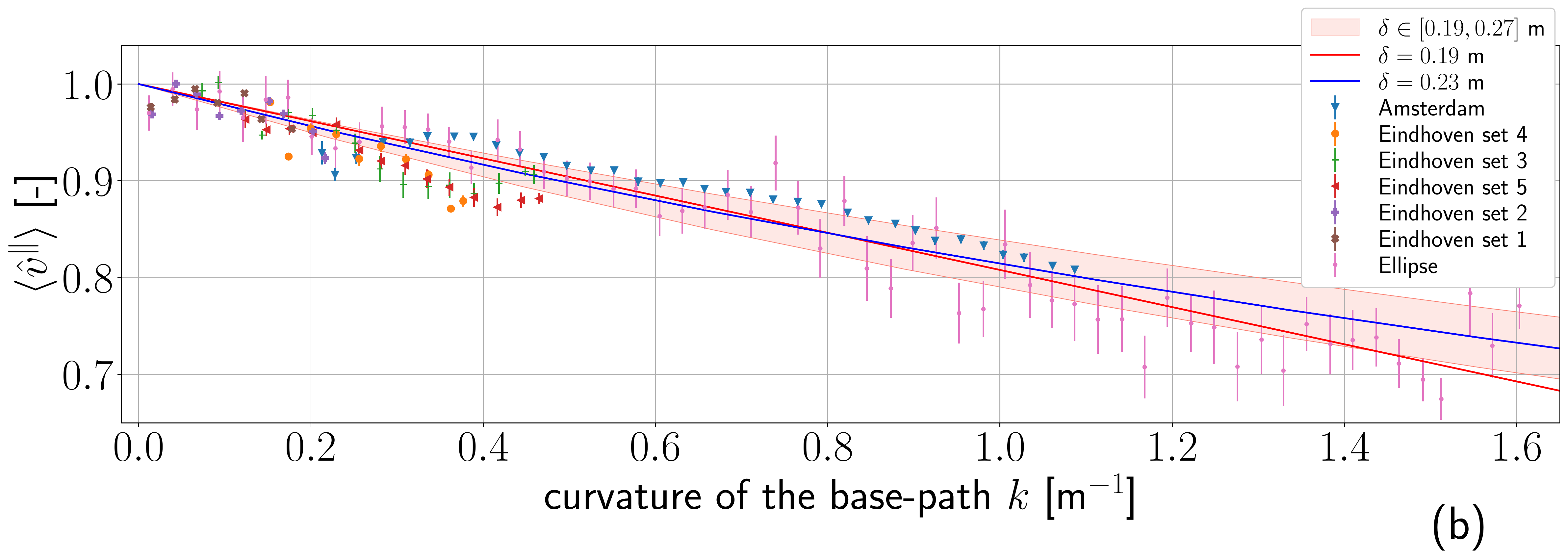}
    \caption{}\label{fig: fundamental diagram}
    \end{subfigure}
    \caption{(a) Sketch of a person following a curved trajectory indicating the body radius, $\delta$, radius of curvature, $R=\frac{1}{k}$, straight-path velocity (SPV), $v_\textsc{\tiny SP}$, and the body center velocity (BCV), $v_\textsc{\tiny BC}$. The linearized velocity of the body center behaves according to Eq.~\eqref{eq:longitudinal velocity-curvature relation}. (b) The average dimensionless longitudinal velocity $\langle \hat{v}^\|\rangle$ as a function of $k$, the curvature of the preferred path for the seven different datasets. The error bars indicate the standard deviation. Fits of Eq.~\eqref{eq: nonlinear fundamental relation} of one hundred random data partitions are represented by the pink area. The results are compared with the fit of Eq.~\eqref{eq: nonlinear fundamental relation} ($\delta=0.23$~m, blue) and the fit of Eq.~\eqref{eq:longitudinal velocity-curvature relation} ($\delta=0.19$~m, red).}
    \label{fig: fundamental diagram and sketch}
\end{figure*}

We assume that body rotation, necessary to adopt a curved trajectory, is the key reason for velocity reduction. Let $\vSP$ denote the velocity pedestrians adopt when walking along straight paths (also Straight-Path Velocity, SPV, henceforth). 
In our datasets $\vSP\in [1.10,1.36]\,$m/s holds, in agreement with literature velocity measurements in the diluted limit (e.g.~\cite{Corbetta2023, Vanumu2017}). 
Suppose a pedestrian with body radius $\delta$ (half body width) walking along a curved path with radius $R=\frac{1}{k}$ (as in Fig.~\ref{fig: sketch}). We expect the velocity of the body parts following the outer bend to remain equal to the straight-path velocity. Assuming a rigid body with shoulder line directed toward the curvature center, the Body Center Velocity (BCV), $v_\textsc{\tiny BC}$, satisfies $v_\textsc{\tiny BC} < v_\textsc{\tiny SP}$, and, the following relation between $v_\textsc{\tiny SP}$, $v_\textsc{\tiny BC}$, $\delta$ and $R$ hold
\begin{equation}\label{eq: nonlinear fundamental relation}
    \frac{v_{\textsc{\tiny BC}}}{R} = \frac{v_{\textsc{\tiny SP}}}{R+\delta}.
\end{equation}
Eq.~\eqref{eq: nonlinear fundamental relation} expresses the physical consequence that under our rigid body and shoulder alignment assumptions, the angular velocity is constant. Linearizing Eq.~\eqref{eq: nonlinear fundamental relation} around $k=0$ returns a more familiar fundamental diagram-like expression 
\begin{equation}\label{eq:longitudinal velocity-curvature relation}
    v_\textsc{\tiny BC}(k)=v_\textsc{\tiny SP}\left(1-k\delta\right).     
\end{equation}
In Fig.~\ref{fig: fundamental diagram} we compare our model with our experimental measurements. We factor out the context-dependency of the velocity, by scaling the BCV to the SPV, i.e.~we consider the following  dimensionless longitudinal velocity at varying curvature
\begin{equation}\label{eq:velocity-curvature-avg}
    \hat{v}^\|(k) = \Bigl\langle{\frac{v_\textsc{\tiny BC}(k)}{v_\textsc{\tiny SP}}}\Bigr\rangle_k,
\end{equation}
where the average is taken among measurements having the same $k$ value (where a binning in $k$ is considered). For each measurement domain, the SPV is determined separately by extrapolating the longitudinal velocity versus curvature relation towards $k=0$.
We report the relation in Eq.~\eqref{eq: nonlinear fundamental relation} with a solid blue line, with body radius $\delta$ fitted to $\delta\approx 0.23$ m. The pink area represents a margin of error obtained by fitting Eq.~\eqref{eq: nonlinear fundamental relation} with 100 random partitions of the data, which are compatible with body radii $\delta\in[19,27]\,$cm consistently with expectations. We report in solid red the linearized relation in Eq.~\eqref{eq:longitudinal velocity-curvature relation} ($\delta=0.19$). Within the curvature range explored ($k\in [0,1.6]\,$m$^{-1}$), the complete (Eq.~\eqref{eq: nonlinear fundamental relation}) and linearized relation (Eq.~\eqref{eq:longitudinal velocity-curvature relation}) appear equally compatible with the data. For technical simplicity, in our Langevin-like model proposed in Section~\ref{section: Langevin-like model for curved tubular neighborhood} we will employ the linearized model.

\begin{figure*}[ht]
    \centering
    \begin{subfigure}{0.49\textwidth}
    \includegraphics[width=\textwidth]{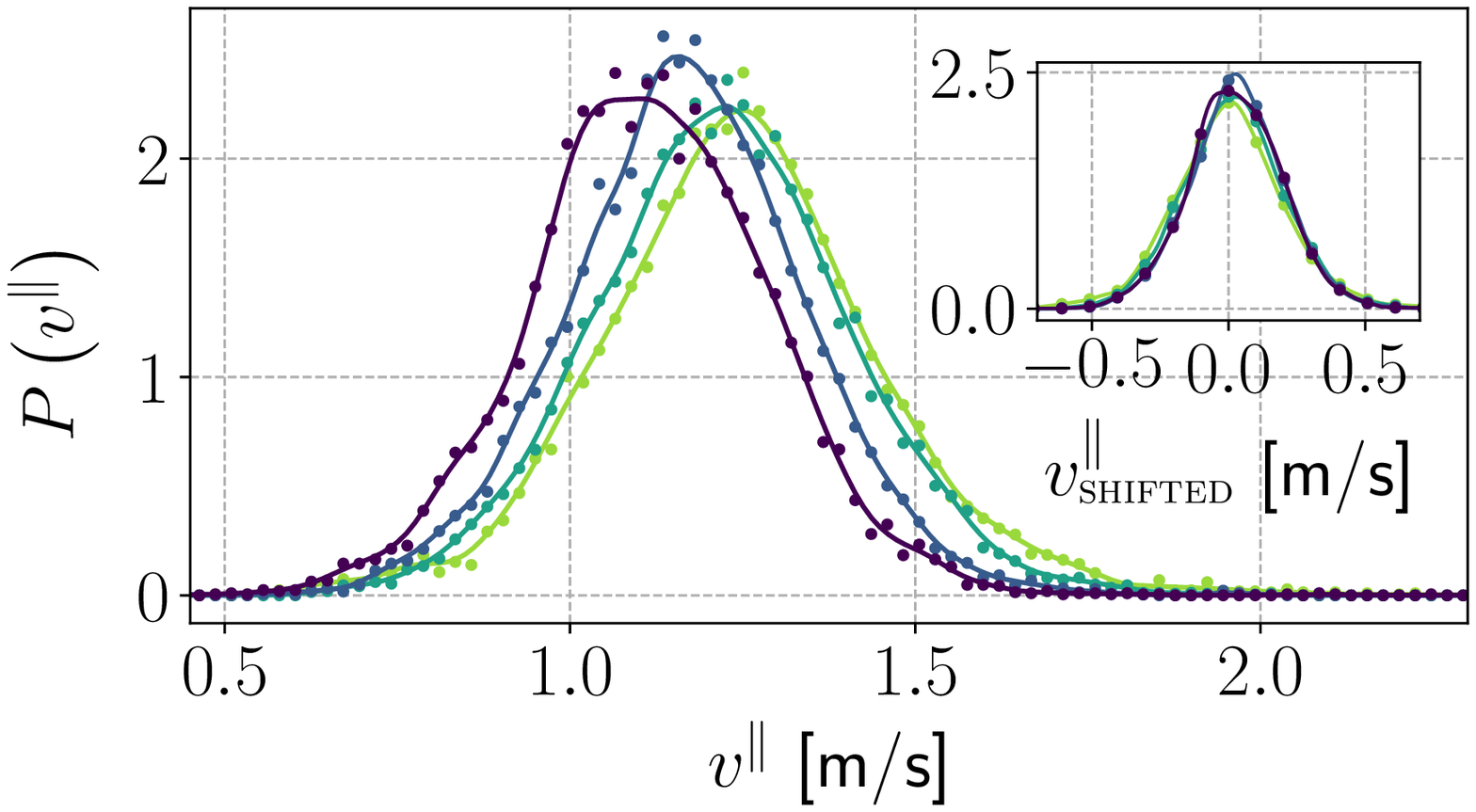}
    \caption{}\label{subfig: longitudinal velocity pdf}
    \end{subfigure}
    \hfill
    \begin{subfigure}{0.49\textwidth}
    \includegraphics[width=\textwidth]{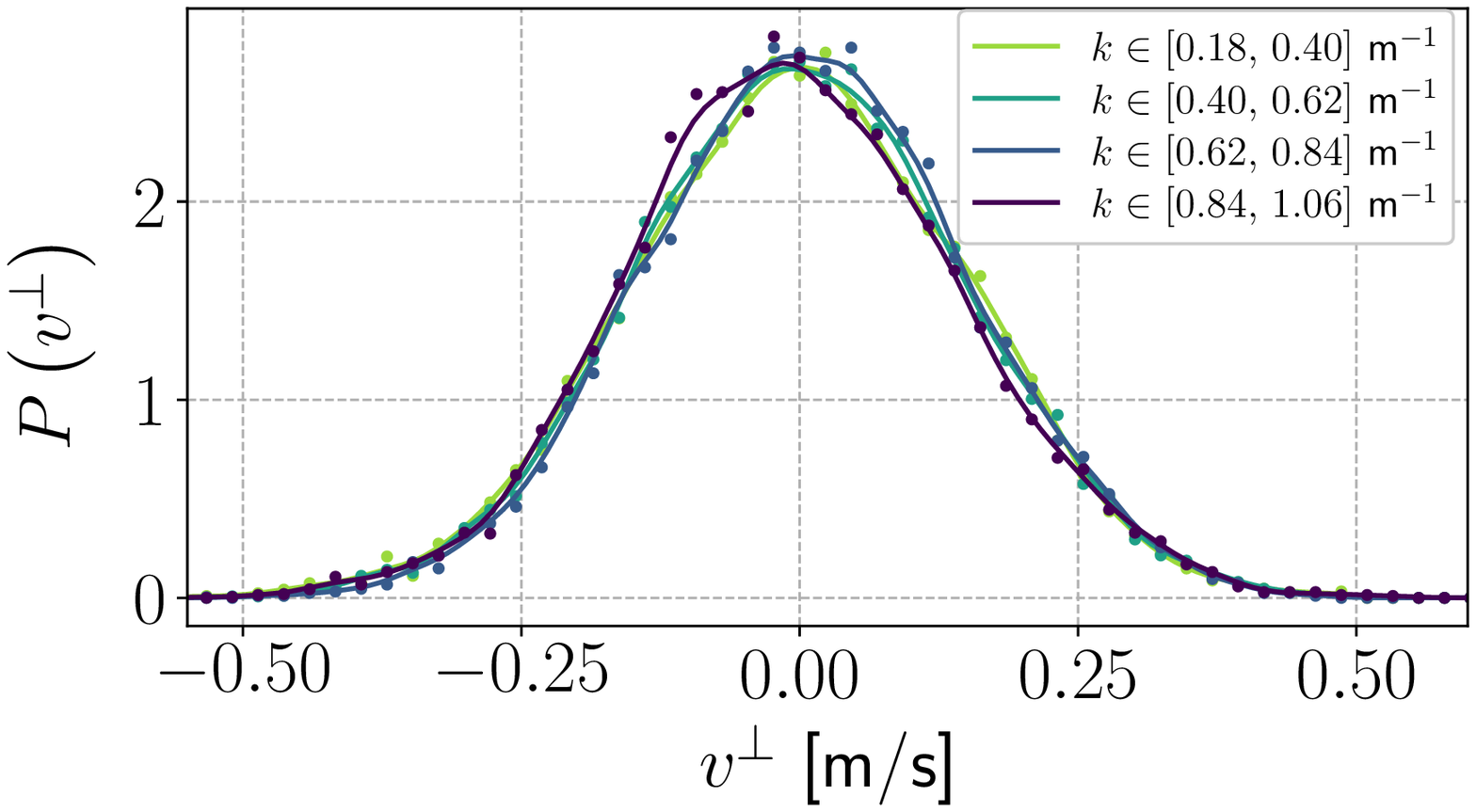}
    \caption{}\label{subfig: transversal velocity pdf}
    \end{subfigure}
    \caption{(a) Probability distribution functions of the longitudinal velocity, $v^\|$, and of the shifted longitudinal velocity, $v^\|_\textsc{\tiny shifted}$, (inset) for four curvature intervals (AMS dataset). The longitudinal velocity fluctuations have different mean. However, the width of the fluctuations is similar for all curvature levels as the inset figure shows. (b) Probability distribution function of the transversal velocity, $v^\perp$, for the same four curvature intervals. Transversal velocity fluctuations are indistinguishable across different curvature levels.}
    \label{fig:pdfs_tangvelo_and_transvelo}
\end{figure*}

\textbf{Velocity fluctuations.} We conclude this section reporting on fluctuations beside the curvature-dependent averages (Eq.~\eqref{eq:velocity-curvature-avg}). 
Due to statistics reasons we focus on our richest dataset, AMS. 
In Fig.~\ref{fig:pdfs_tangvelo_and_transvelo} we report the probability density function of longitudinal ($v^\|$) and transversal ($v^\perp$) velocity fluctuations. 
In line with the fundamental diagram (Fig.~\ref{fig: fundamental diagram}), the means of the longitudinal velocity decrease for higher curvature levels. Compensating for this shift considering
\begin{equation}
    v^\|_\textsc{\tiny shifted}:=v^\|-v_\textsc{\tiny BC}(k)
\end{equation}
 with $\delta=19$~cm (cf. Fig.~\ref{subfig: longitudinal velocity pdf}), it can be seen that fluctuations in the (shifted) longitudinal velocity are curvature independent and have a Gaussian fluctuation structure with standard deviation $\sigma_{v^\|} = 0.19$~m/s. 
 Similarly, fluctuations in transversal velocity do not depend on the curvature and have Gaussian fluctuation, $\sigma_{v^\perp} = 0.15$~m/s. These measurements, after velocity shifts are compatible with experimental campaigns focusing on straight paths~\cite{Corbetta2017,Corbetta2018}. Similarly to~\cite{Corbetta2017,Corbetta2018}, the Gaussian behavior of velocity fluctuations will be crucial in modeling perspective, posing the bases to our Langevin-like structure.

\section{Langevin-like model for curved tubular neighborhood}\label{section: Langevin-like model for curved tubular neighborhood}
In this section we show that the walking dynamics around preferred paths can be modeled quantitatively with a Langevin-like model defined on the tubular neighborhood of $\boldxgamma$. 

 \textbf{Fluctuations around straight paths.} The model introduced here extends the Langevin-like model previously proposed by some  of the authors and that addresses the case in which $\boldxgamma$ is a straight trajectory \cite{Corbetta2017}. In~\cite{Corbetta2017}, the fluctuating motions of pedestrians have been modeled as a superposition of social forces determining the individual acceleration, $\ddot{\boldsymbol{x}}$. Assuming for simplicity a coordinate system $(x,\,y)$ in which $\boldxgamma$ is the path $y=0$, thus $x$ identifies the position along $\boldxgamma$, and $y$ is the transversal coordinate (i.e.~$\boldxgamma = (s,\,0)$, $(\eunit_\|,\,\eunit_\perp) \equiv(\eunit_x,\,\eunit_y)$, $(\dot x,\,\dot y) = (v^\|,\,v^\perp)$), individual accelerations read
\begin{equation}\label{eq: eq of motion straight paths}
\begin{aligned}
 \ddot{\boldsymbol x} &= f\left(\dot x,\vBC\right)\mathbf{e}_x + \left(-2 \beta y -2\mu \dot{y}\right)\mathbf{e}_y + \sigma\dot{\boldsymbol{W}}\\
 &= f\left(v^\|,\vBC \right)\mathbf{e}_\| + \left(-2 \beta y -2\mu v^\perp\right)\mathbf{e}_\perp + \sigma\dot{\boldsymbol{W}}.
\end{aligned}
\end{equation}
The previous equation models the following effects
\begin{itemize}
    \item[E1 -] self-propulsion along $\boldxgamma$ driven by $f\left(v^\|\right)$. At first-order Taylor expansion $f\left(v^\|,\,\vBC\right)$ is a relaxation term towards a desired walking speed for the body center $\vBC$, i.e.
    \begin{equation}\label{eq:fvvbc}
        f\left(v^\|,\,\vBC\right) = -2\alpha\left(v^\|- \vBC\right),
    \end{equation}
    where $\alpha$ is inversely proportional to the time-scale $\tau = (2\alpha)^{-1}$ for relaxation towards the desired velocity. Note that this term can be interpreted as an active viscous term with quadratic velocity potential 
    \begin{equation}
        \Phi_\|\left(v^\|,\,\vBC\right) = \alpha\left(v^\| - \vBC\right)^2.
    \end{equation}
    \item[E2 -] transversal confinement in the $\boldxgamma$ neighborhood, and transversal velocity damping, which is modeled as a damped harmonic oscillator. This is parametrized by a linear stiffness coefficient $\beta$ and a linear friction coefficient $\mu$. 
    \item[E3 -] random noise, $\dot {\boldsymbol W} := (\dot{W}^\|,\,\dot{W}^\perp)$, to generate fluctuations and recover randomness in behavior. For simplicity, this is assumed to be $\delta$-correlated in time, isotropic, with components mutually uncorrelated Gaussian distributed ($\sigma$ is a scale parameter). This hypothesis quantitatively agrees with the observed fluctuations in terms of correlation structure and probability density of velocities and positions.
\end{itemize}
Note that in~\cite{Corbetta2018}, pair-wise interactions to reproduce the statistics of the avoidance behavior have been included in this model.

\textbf{Parallel dynamics in a tubular neighborhood: geometric setting.} 
Here we extend model in Eq.~\eqref{eq: eq of motion straight paths} to include curvature effect.
When pedestrians follow a path with small curvature, we do not expect effects due to curvature: path appears locally straight. Pedestrians in these conditions would walk following their curved, preferred path. We incorporate this aspect in the left-hand-side of the equation of motion~\eqref{eq: eq of motion straight paths}. Heuristically, we opt to vary the underlying geometry.

\begin{figure}[H]
    \centering
    \includegraphics[trim=3cm 4.5cm 2.2cm 1.1cm,clip,width=0.8\columnwidth]{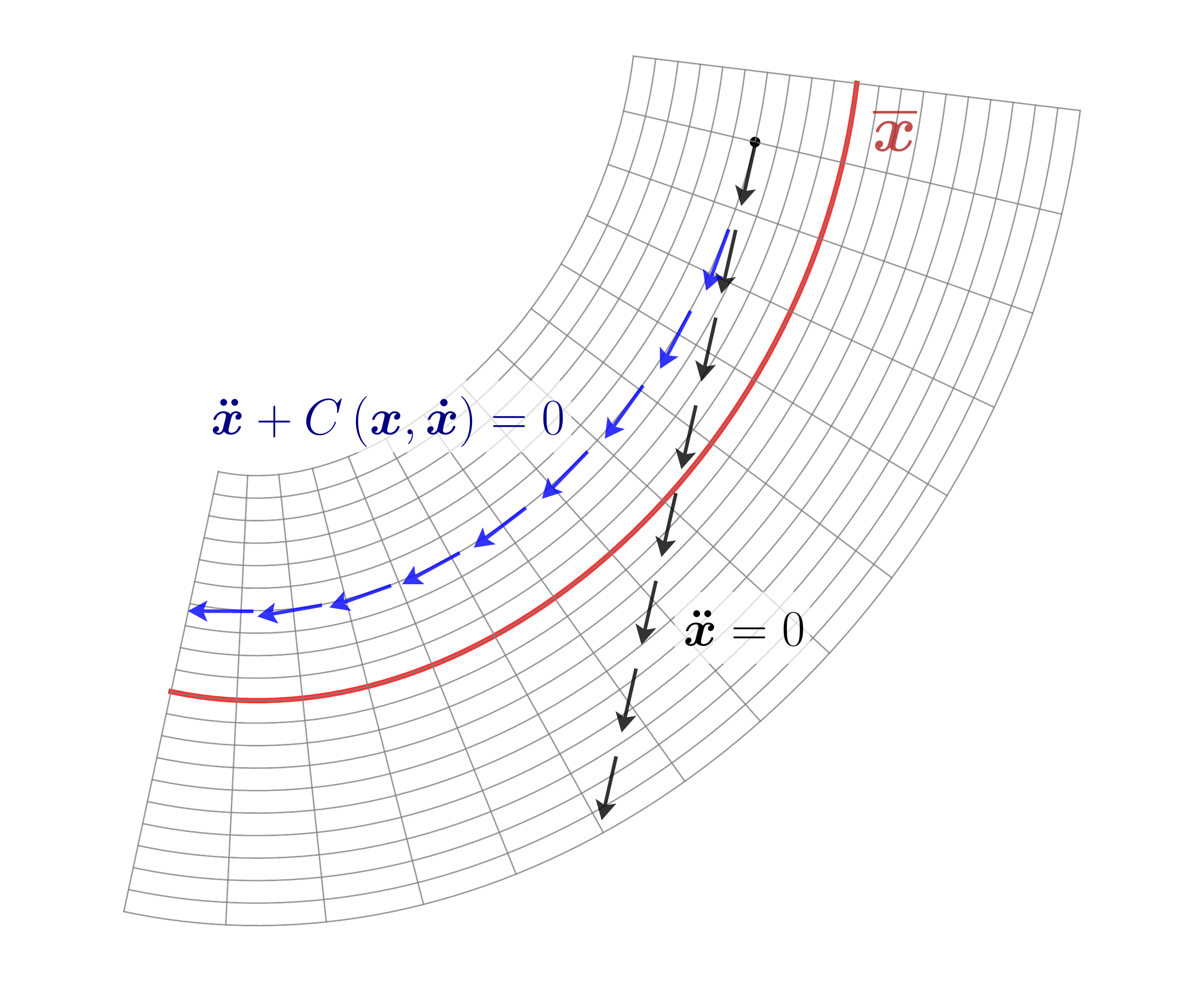}
    \caption{A tubular coordinate frame around $\boldxbar$ with the evolution of a velocity vector respecting $\Ddot{\boldsymbol x}=0$ and $\nablauu = \Ddot{\boldsymbol x} + C\left(\boldsymbol x, \dot{\boldsymbol x}\right)=0$ in black and blue, respectively.}
    \label{fig:sketch parallel transport}
\end{figure}

First, in absence of forces and noise, Eq.~\eqref{eq: eq of motion straight paths} describes a pedestrian conserving their initial momentum:
\begin{equation}\label{eq: ddotx=0}
\ddot {\boldsymbol x} = \boldsymbol 0 \quad\Rightarrow\quad \dot {\boldsymbol x} = const.
\end{equation}
This translates into a rectilinear motion (depicted by the black arrows in Fig.~\ref{fig:sketch parallel transport}).

We generalize the left-hand-side of Eq.~\eqref{eq: ddotx=0}, considering broader possibilities of force-free curves (typically addressed as geodesic curves) as solutions of
\begin{equation}\label{eq:geodesic flow}
    \nablauu := \Ddot{\boldsymbol x} + C\left(\boldsymbol x, \dot{\boldsymbol x}, \boldxbar\right) = 0.
\end{equation}
Here, we adopt the notation $\nablauu$ for the covariant derivative of $\dot{\boldsymbol x}$. In the field of differential geometry, the covariant derivative is commonly used to express the change of vectors when transporting them in a (curved) geometry~\cite{jost2008}. 
Additionally, the correction term, $C\left(\boldsymbol x, \dot{\boldsymbol x}, \boldxbar\right)$, is usually expressed by so-called Christoffel symbols of the second kind: $C\left(\boldsymbol x, \dot{\boldsymbol x}, \boldxgamma\right) := \sum_{i,j,k=1,2}\Gamma^i_{k j}\dot{x}^k\dot{x}^j\eunit_i$ (where the indexed notation satisfies $(x^1,x^2):=(x,y)$, $(\mathbf{e}_1,\mathbf{e}_2):= (\mathbf{e}_x,\mathbf{e}_y)$).
Technical properties of the covariant derivative and Christoffel symbols are postponed to Appendix~\ref{appendix:diffgeo}.

Our tailored correction term is constructed such that geodesic curves (i.e.~solutions of Eq.~\eqref{eq:geodesic flow}, cf. example blue arrows in Fig.~\ref{fig:sketch parallel transport}) respect the following physical properties:
\begin{itemize}
    \item geodesic curves conserve the (Euclidean) kinetic energy, i.e.
    \begin{equation}
     \nablauu = 0 \Rightarrow \frac{d}{dt}\|\dot{ \boldsymbol { x}}\|^2 =   \frac{d}{dt} \left(\dot{x}^2+\dot{y}^2\right) = 0.
    \end{equation}
    \item geodesic curves  initially parallel to $\boldxgamma$, i.e.~with zero initial orthogonal velocity, remain parallel to $\boldxgamma$ at all times. In formulas
    \begin{equation}
    \begin{cases}
        v^\perp(t = 0) = 0\\
        \nablauu = 0
    \end{cases}\Rightarrow 
        v^\perp(t) = 0,\ \forall t > 0
    \end{equation}
    or, equivalently, in components
    \begin{equation}
    \begin{cases}
       \dot h(t=0) = 0\\
       \nablauu = 0
     \end{cases}\Rightarrow \dot h(t) = 0,\  \forall t > 0.
    \end{equation}
\end{itemize}
This means that if $\boldxgamma$ is not straight, also geodesics will not be.  Two examples of geodesics are shown in Fig.~\ref{fig: sketch geodesic flow} (curve $\mathbf{a}$ and $\mathbf{b}$). It can be seen that the properties of remaining parallel and conserving mechanical energy are satisfied, which is ensured by the centripetal-like acceleration $C\left(\boldx, \dot{\boldx}, \boldxgamma\right)$ - (note that this is not a covariant derivative of Levi-Civita type for the Euclidean metric). In our forthcoming simulations we opt to generate trajectories in the physical $(x,y)$ coordinates. This allows to easily account for forcing terms and possibly generalize our work to include interactions. 
On the other hand, the correction term remain defined via an implicit system of equations. 
To prevent this section from becoming needlessly technical, we opt to postpone our derivation of the expression of the correction term following the two hypotheses above as well as their transformation in $(x,y)$ coordinates in Appendix~\ref{appendix:diffgeo}.

\textbf{Pedestrian fluctuations in a tubular neighborhood.} To model the fluctuating behavior of pedestrians walking along curved paths, we perturb the force-free dynamics described by Eq.~\eqref{eq:geodesic flow}, including counterparts of the effects E1-E3. We additionally hypothesize, consistently with the fundamental diagram in Sect.~\ref{section:curvature velocity fundamental diagram and fluctuations}, that the body center velocity depends on the instantaneous curvature following Eq.~\eqref{eq:longitudinal velocity-curvature relation}.
We assume that pedestrians (in absence of stochastic fluctuations) can adjust instantaneously to such velocity as the curvature changes along the path (i.e.~when $\dot k \neq 0$). In other terms, the combination of Eq.~\eqref{eq:fvvbc} and Eq.~\eqref{eq:longitudinal velocity-curvature relation} would provide a curvature dependent propulsion force $f(v^\|,\vBC(k))$. Yet, a propulsion force $f(v^\|,\vBC(k))$ built by bare combination of the two terms, would take a time $\tau >0$ to relax to changes in desired velocity due to changed curvatures. We instead assume that pedestrians are instantaneously capable of adjusting to variations in curvature. This can be modeled by correcting the propulsion term including a contribution of the curvature time gradient $\dot k$. This yields a corrected propulsion $\hat f(v^\|,\vBC(k),\dot k)$ which reads
\begin{eqnarray}
    \hat f(v^\|,\vBC(k),\dot k) &=& f(v^\|,\vBC(k)) - \vSP\delta\dot k\label{eq:correctedfhat}\\ 
    &=& -2\alpha\left(v^\| -\vSP (1-\delta k)+\frac{\vSP\delta}{2\alpha}\dot{k}\right). \nonumber
\end{eqnarray}
Note that $\hat f \equiv f$ whenever the curvature gradient is zero, e.g.~on straight paths.

Combining our geodesic flow parallel to the curved preferred path $\boldxgamma$ (Eq.~\eqref{eq:geodesic flow}), the effects E1-E3, abd the corrected propulsion term in Eq.~\eqref{eq:correctedfhat} yields the following force balance
\begin{equation}\label{eq:main}
    \nabla_{\Dot{\boldsymbol x}}\Dot{\boldsymbol  x} = \hat f(v^\|)\mathbf{e}_\| + \left(-2 \beta h -2\mu v^\perp\right)\mathbf{e}_\perp +  \sigma\dot{\boldsymbol{W}}.
\end{equation}
An example of a trajectory generated by this model is in Fig.~\ref{fig: sketch geodesic flow} (curve $\mathbf{c}$), where the modeling forces confine the trajectory around the preferred path $\boldxgamma$. In the next section we show that Eq.~\eqref{eq:main} describes quantitatively the statistics of the fluctuations of pedestrians walking about  curved paths.

\begin{table}[H]

\caption{Estimated parameters of the model for the Amsterdam train station dataset. 
$\alpha$, modulating factor of longitudinal propulsion force; 
$\beta$, stiffness coefficient of the transversal linear Langevin
dynamics; 
$\mu$, friction coefficient of the transversal linear Langevin
dynamics; 
$\sigma$, white noise intensity;
$\vSP$, straight-path velocity;
$\delta$, body radius. The parameter estimates and are obtained by fittings of the fundamental diagram, the $\vshift$-time correlation and Langevin potentials. Further details on the parameter estimation, including error analysis, are provided in Appendix~\ref{Appendix: model calibration}.}
\begin{center}
\begin{tabular}{l|l|l}
\hline
parameter             & value & \\ \hline
$\alpha$              & 0.26  & s$^{-1}$\\
$\beta$               & 1.17   & s$^{-2}$\\
$\mu$                 & 0.39  & s$^{-1}$\\
$\sigma$              & 0.19  & ms$^{-3/2}$\\
$v_\textsc{\tiny{SP}}$       & 1.33 & ms$^{-1}$\\
$\delta$              & 0.192 & m\\ \hline
\end{tabular}\label{table: parameters model AMS}
\end{center}
\end{table}

\begin{figure*}[ht]
    \centering
    \begin{subfigure}{0.47\textwidth}
    \includegraphics[width=\textwidth]{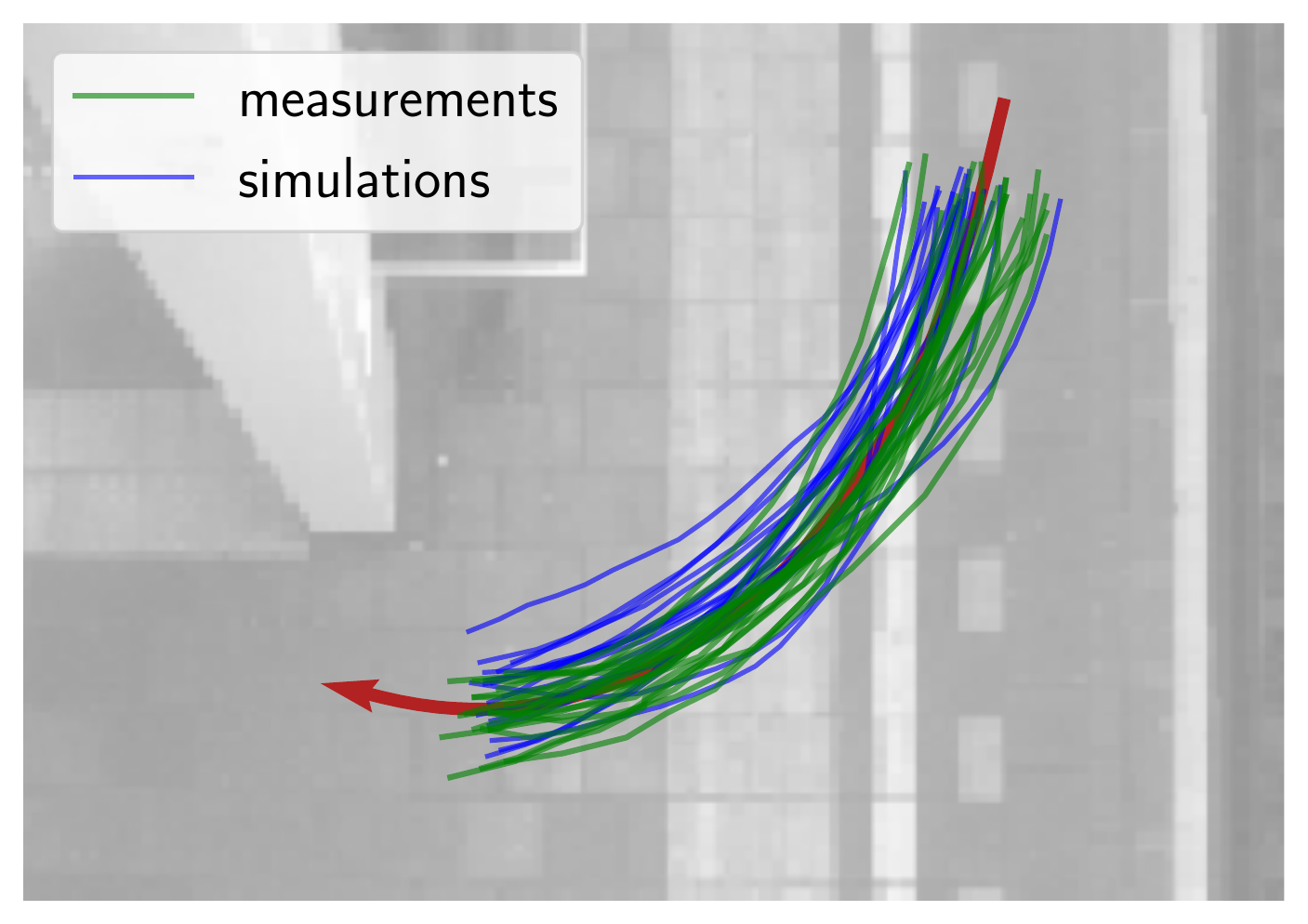}
    \hspace{3cm}\caption{}
    \label{fig: top view simulated trajectories AMS}
    \end{subfigure}
    \hfill
    \begin{subfigure}{0.47\textwidth}
    \includegraphics[width=\textwidth]{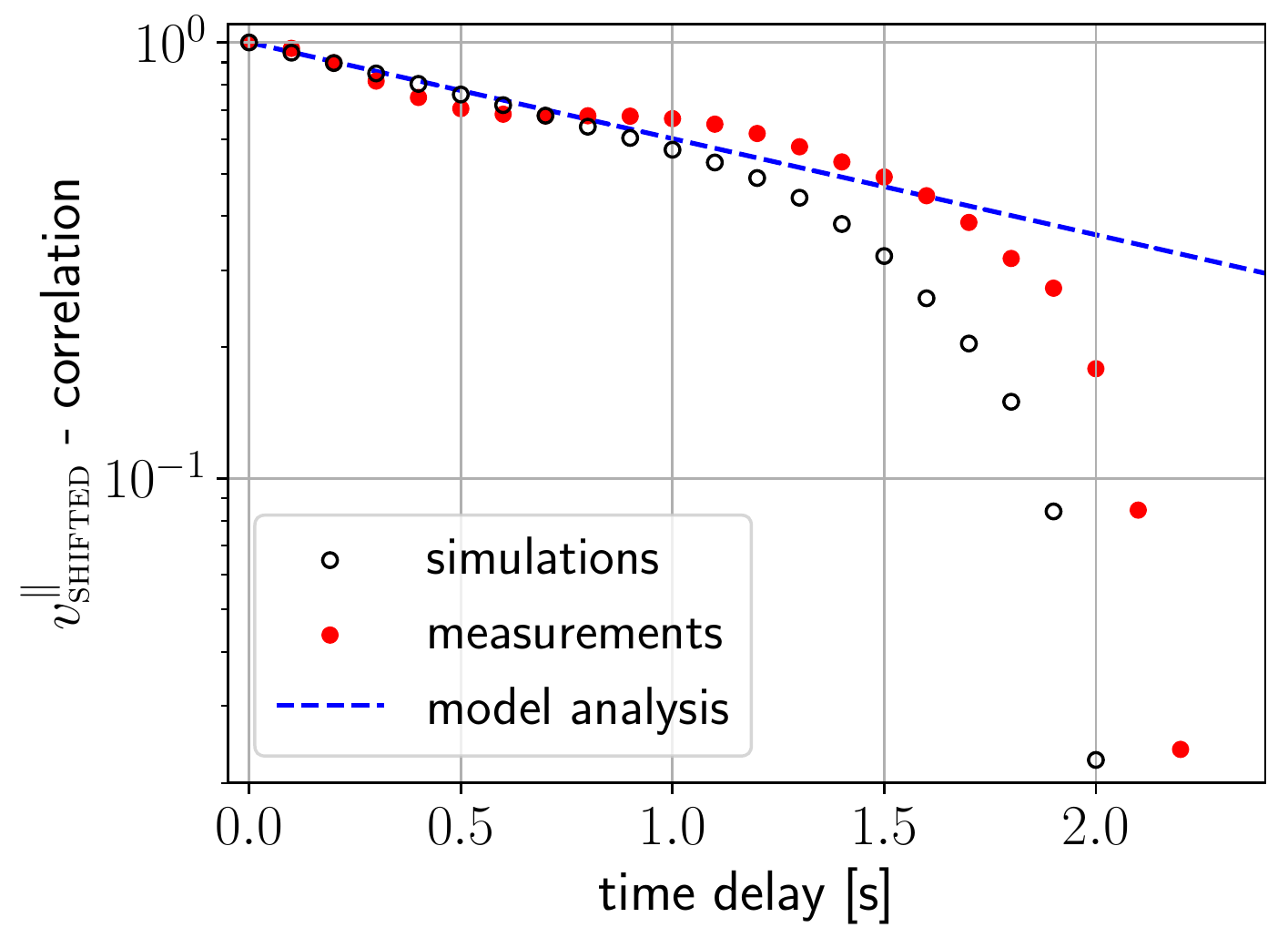}
    \caption{}
    \label{fig: V shifted correlation AMS}
    \end{subfigure} 
    
    \begin{subfigure}{0.325\textwidth}
    \includegraphics[width=\textwidth]{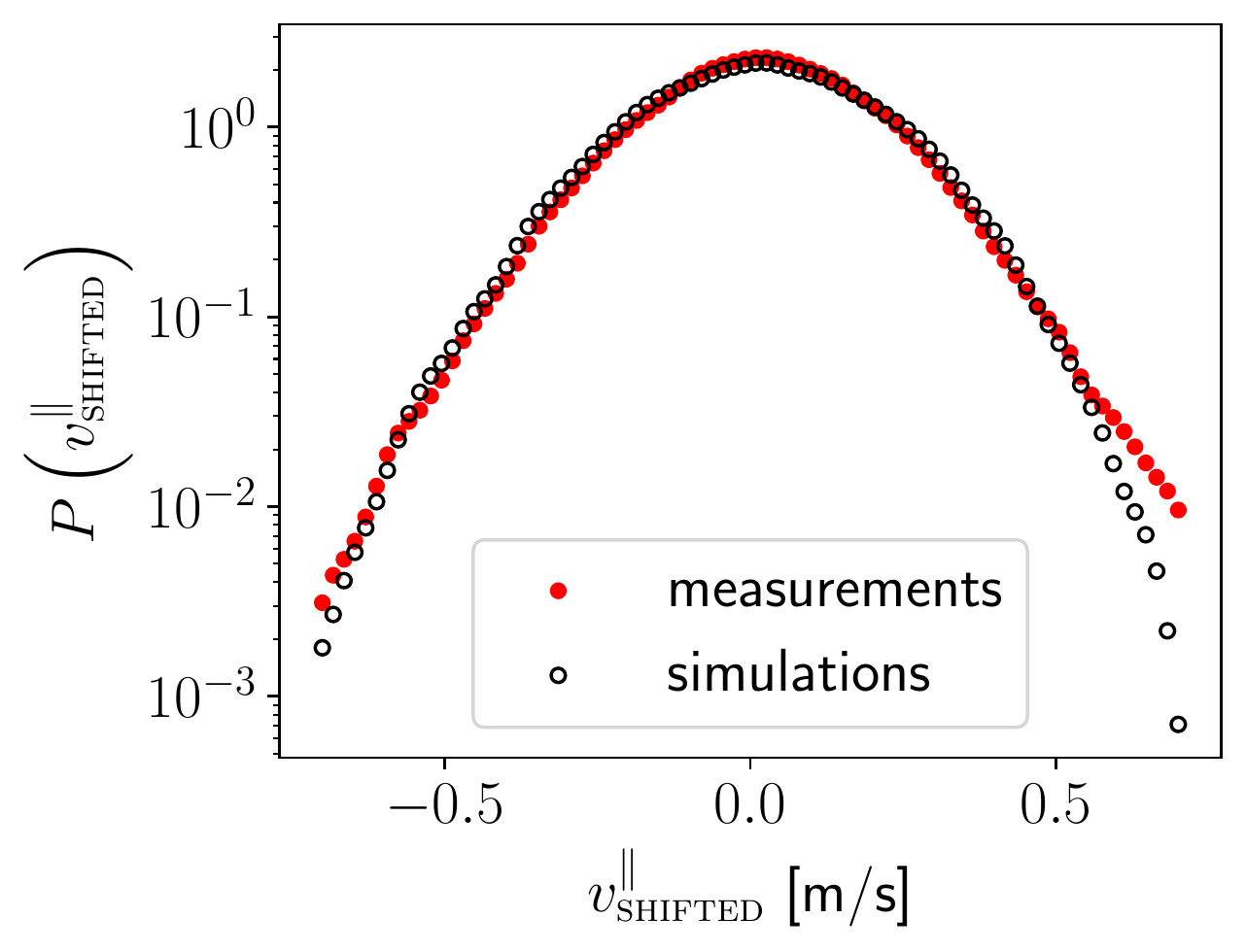}  
    \caption{}\label{fig: pdf comparison V longitudinal shifted AMS}
    \end{subfigure}
    \begin{subfigure}{0.325\textwidth}
    \centering
    \includegraphics[width=\textwidth]{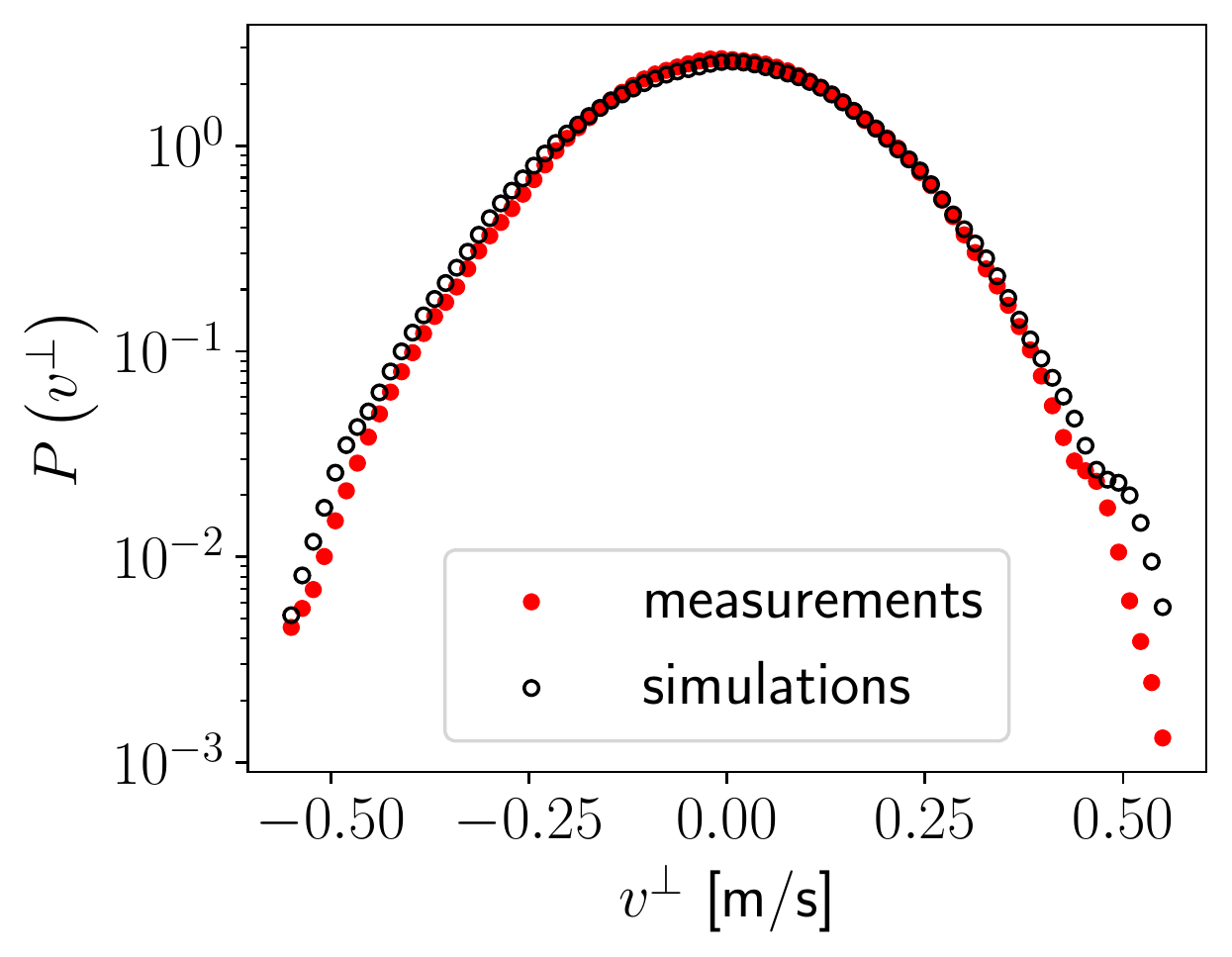}  
    \caption{}
    \label{fig: pdf comparison V perp AMS}
    \end{subfigure}
    \begin{subfigure}{0.325\textwidth}
    \centering
    \includegraphics[width=\textwidth]{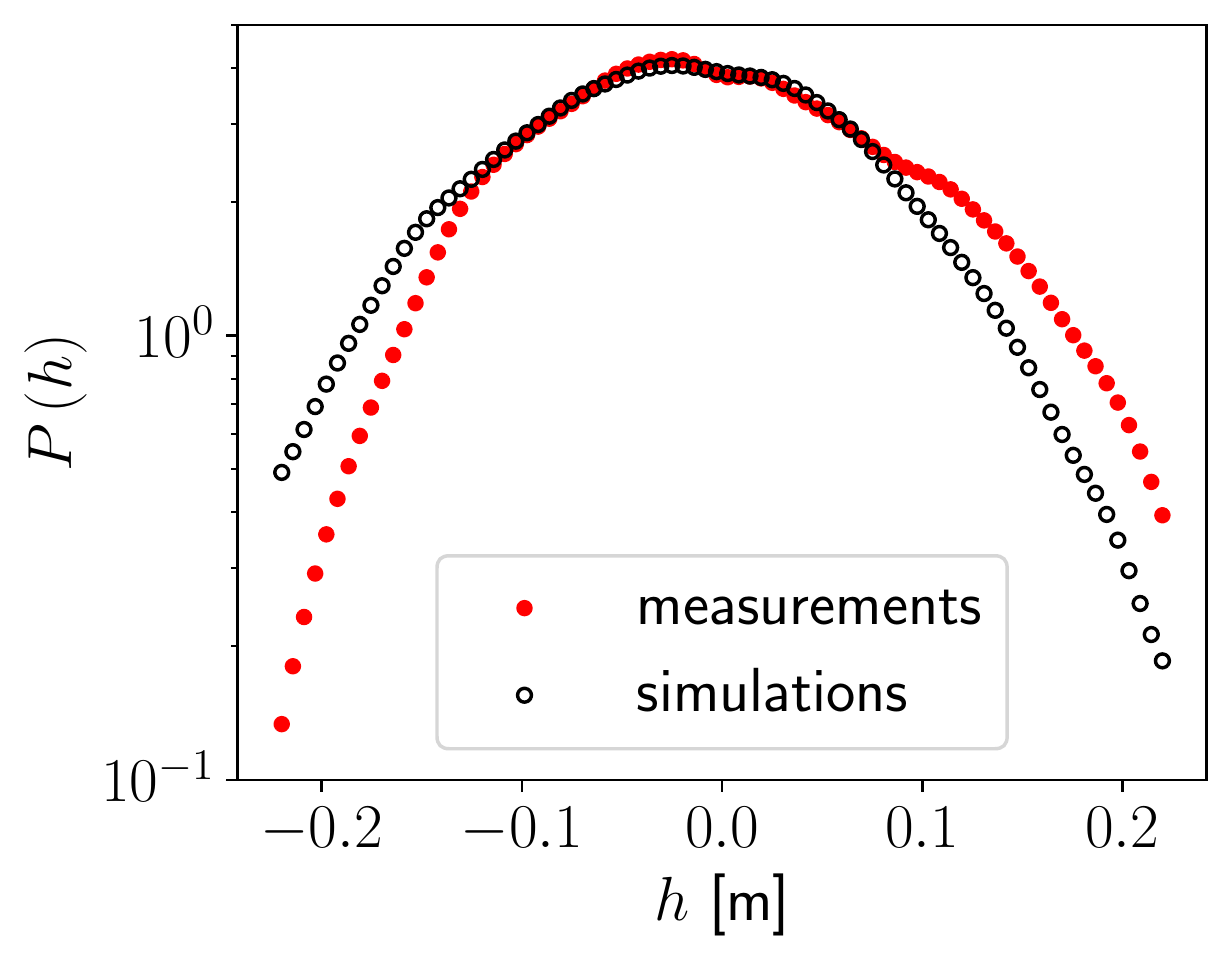}  
    \caption{}
    \label{fig: pdf comparison h AMS}
    \end{subfigure}
    
    \caption{(a) Top view of Amsterdam south train station platform 2.1 with average path, measured trajectories and simulated trajectories. (b) Time correlation of the shifted longitudinal velocity, $v^\|_\textsc{\tiny shifted}$, of the measured trajectories in Amsterdam train station (red). The fitted analytic exponential decay [$\exp{\left(-2\alpha t\right)}$] - blue dotted line - is compared with the measurements and simulations. (c-e) Comparison of the empirical probability distribution functions of the shifted longitudinal velocity (c), transversal velocity (d) and lateral deviation (e) with the distribution functions of the simulation in Amsterdam train station.}
    \label{fig:my_label}
\end{figure*}

\section{Results}\label{section:resulst}
In this section we compare the stochastic dynamics modeled by Eq.~\eqref{eq:main} with experimental data. We focus on trajectories following the curved path at Amsterdam South station (AMS dataset), as it is the richest in amount of trajectories allowing to fully resolve and compare statistical fluctuations. We consider the SPV, $\vSP$, and body size radius $\delta$ determined in Appendix~\ref{Appendix: model calibration}. We estimate the scale parameters ($\alpha$, $\beta$, $\mu$, $\sigma$) by considering Langevin potentials in longitudinal velocity (shifted as in Sect.~\ref{section:curvature velocity fundamental diagram and fluctuations})
\begin{equation}
\Phi_{v^\|_\textsc{\tiny shifted}}\sim -\log \Prob (v^\|_\textsc{\tiny shifted}),
\end{equation}
$\Prob(v^\|_\textsc{\tiny shifted})$ here indicates the probability density of $v^\|_\textsc{\tiny shifted}$, and lateral deviation $\Phi_{h}$  and transversal velocity $\Phi_{v^\perp}$. The fitting procedure follows the approach in~\cite{Corbetta2017,Corbetta2018}, and technical details are in Appendix~\ref{Appendix: model calibration}. We report the values of the model parameters in \autoref{table: parameters model AMS}.

With the estimated parameters from \autoref{table: parameters model AMS} and the simulation procedure explained in Appendix~\ref{Appendix: numerical simulations}, we perform simulations of 2,700 trajectories with a discretization step size of 0.1 seconds, comparable to dataset AMS. Fig.~\ref{fig: top view simulated trajectories AMS} displays a collection of simulated trajectories, qualitatively indistinguishable from the measurements.

Next, we consider stochastic properties by comparing the empiric and simulated probability distributions of the fluctuations in three observables: shifted longitudinal velocity, transversal velocity and lateral deviation. The empiric probability distribution functions, as well as the ones obtained from the simulations, are shown in Fig.~\ref{fig: pdf comparison V longitudinal shifted AMS}-\ref{fig: pdf comparison h AMS}. It can be seen that the stochastic properties of the velocity fluctuations are captured by the model. The simulated fluctuations in transversal position are also in good agreement. 
However, for lateral deviations larger than $10$~cm ($|h|>0.1$~m), we observe that the empirical fluctuations deviate from the Gaussian behavior.
This could potentially be attributed to architectural constraints within the station (e.g.~the entrance of the staircase) which could impede inward ($h<0$) and facilitate outward fluctuations ($h>0$).

Another important statistical property, also used in the model calibration, is the correlation of the shifted longitudinal velocity. In Fig.~\ref{fig: V shifted correlation AMS}, it can be seen that the empiric $v_\textsc{\tiny shifted}^\|$-correlation is recovered reasonably well by the model.

\section{Discussion}
We have investigated the fluctuating dynamics of undisturbed pedestrians walking along curved paths with high statistical,  space- and time-accuracy. Our analysis hinged on large trajectory datasets acquired in both real-life conditions and in a experimental set-up. The trajectories in the datasets cover a broad range of curvature radii. Thanks to these, we have shown that in the diluted limit a fundamental diagram-like relation between the average longitudinal walking velocity and path curvature exists. Specifically, the average longitudinal velocity decreases for increasing curvature. Notably this reduction is quantitatively compatible with a basic rigid-body-like kinematic model. A first-order expansion of such a model, yield a fundamental diagram-like relation. Based on the large datasets, we have analyzed pedestrian motion beyond averages targeting  fluctuations in velocity and lateral deviation. These display Gaussian statistics. Besides, the amplitude of the velocity fluctuations (variance) is independent on the curvature level, at for the range of curvatures observed ($k\in [0,1]\,$m$^{-1}$).

Based on these findings, we have extended the quantitative Langevin-like model by Corbetta \textit{et al.}~\cite{Corbetta2017} to reproduce, in a statistically quantitative way, the walking dynamics of pedestrians along generic, curved, average paths. In our model, we have considered pedestrians as particle moving according to a custom geodesic flow shaped after the average path. The geodesics we consider are characterized by the conservation of kinetic energy and by the fact that they remain parallel to the average path (when the initial velocity is). 
We have modeled pedestrian dynamics by perturbing this geodesic flow by (social-like) forces representing (lateral) path adherence, longitudinal propulsion, and random noisy fluctuations. 
We have validated the model by comparing the probability density functions and the correlation functions generated by repeated model simulations with our measurements at Amsterdam South station. Our model successfully captures the stochastic features of the motion in terms of fluctuations in velocity and position.

We have opted to operate in Cartesian coordinates within a curved geometry, embedding curvature effects in a custom covariant derivative. We believe this choice is instrumental towards further generalization of the model to include, e.g., interactions with other pedestrians and/or different types of forces or noise. All these are typically addressed in Cartesian coordinates. Within the geometric framework we propose, in fact, no coordinate transformations of the forces are required, but only a computation of a correction term (i.e. a Christoffel symbol). Mapping interaction forces in the local coordinate system of each pedestrian would rapidly turn prohibitively complex and computationally expensive.

\appendix 

\section{Definition of preferred path $\boldxgamma$}\label{Appendix: average}
For the definition of the preferred path, we consider a bundle with $N$ trajectories, 
\begin{equation*}
    \{\boldx_\nu(t)\mid \nu=1,2,\ldots,N\}. 
\end{equation*}
Due to the variability in the velocity of pedestrians, we parameterize each trajectory by the relative time, 
\begin{equation*}
s := \frac{t-t_1}{t_2-t_1},
\end{equation*}
where $t_1$ and $t_2$ are the times that a trajectory enters and leaves the measurement site, respectively.
The preferred path, $\boldxgamma$, is defined as an ensemble average over the bundle at each relative time instance $s\in[0,1]$:
\begin{equation}\label{eq: average path equation}
\boldxgamma =\langle \boldx_\nu(s)\rangle_\nu = \frac{1}{N}\sum_{\nu=1}^N\boldx_\nu(s).
\end{equation}

\section{Data selection procedure}\label{Appendix: data selection}

\textbf{Trajectory selection AMS.} To ensure that the data only contains trajectories under diluted conditions, we restrict to trajectories tracked when no other pedestrian is tracked on the platform. We, furthermore, restrict to walking-speed by removing trajectories with average velocity outside $[0.5, 2.5]$ ms$^{-1}$.

\begin{figure*}[ht]
    \centering
    \begin{subfigure}{0.347\textwidth}
    \includegraphics[width=\textwidth]{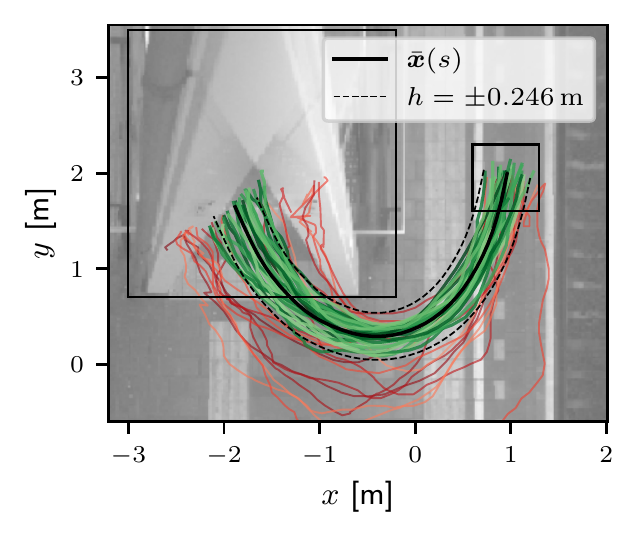}
    \caption{}\label{fig: trajectory selection Amsterdam}
    \end{subfigure}
    \hfill
    \begin{subfigure}{0.321\textwidth}
    \includegraphics[width=\textwidth]{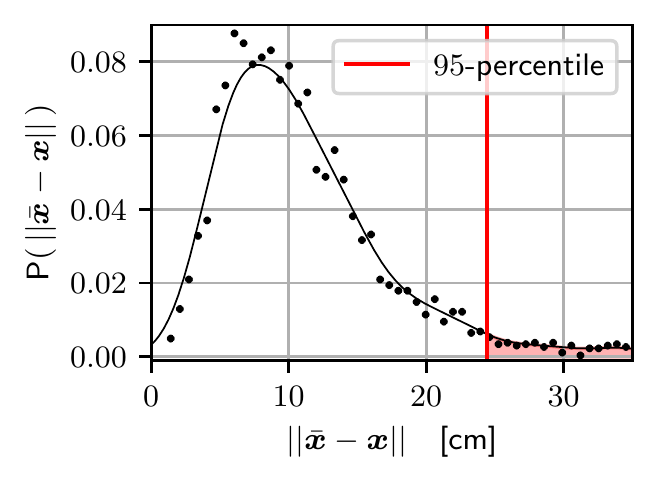}
    \caption{}\label{fig:data selection AMS deviation}
    \end{subfigure}
    \hfill
    \begin{subfigure}{0.321\textwidth}
    \includegraphics[width=\textwidth]{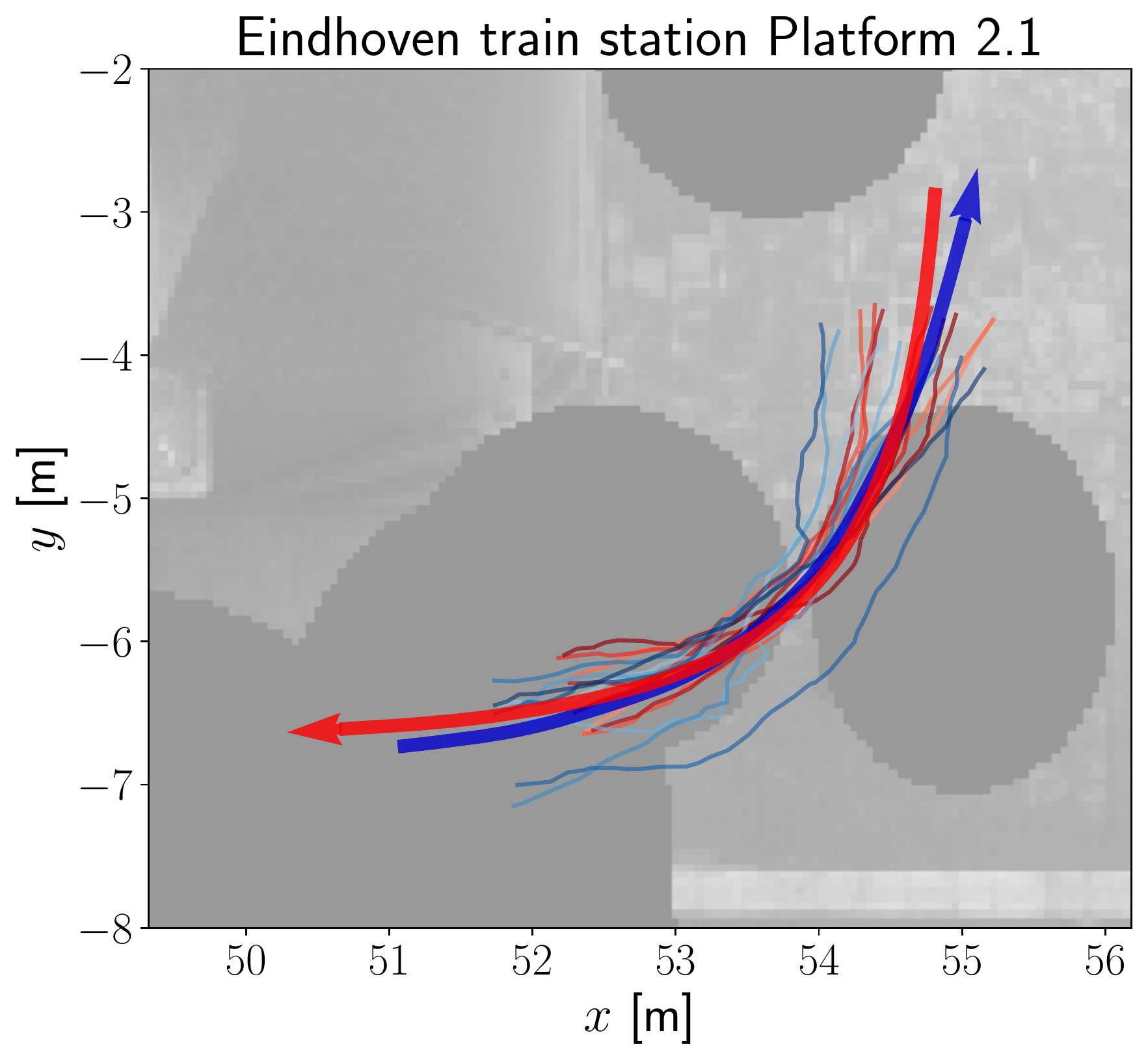}
    \caption{}\label{fig:remaining sets ehv}
    \end{subfigure}
    \caption{(a) Collection of trajectories at Amsterdam south train station platform 2.1 including the origination and termination rectangles. Trajectories among the 95\% least deviating trajectories are colored green. The 95-percentile of average absolute deviation ($\lVert\boldxbar - \boldx\rVert = 24.6\,\mathrm{cm}$) is indicated with the dotted line. The green trajectories are selected. (b) Probability distribution function of the trajectory deviation from the preferred path ($\lVert\boldxbar - \boldx\rVert$). The 95-percentile, indicated with the red line, is used for trajectory selection. (c) The average paths of measurement bundle 4 (blue) and bundle 5 (red) with some selected trajectories in Eindhoven train station.}
    \label{fig: data selection}
\end{figure*}

We consider the bundle with trajectories starting near the railroad 
(i.e.~$(x,y) \in [0.6, 1.3] \times [1.6, 2.3] \, \mathrm{m}^2$)
and finishing at the staircase 
(i.e.~$(x,y) \in [-3.0, -0.2] \times [0.7, 3.5] \, \mathrm{m}^2$) 
depicted by the two rectangles in Fig.~\ref{fig: trajectory selection Amsterdam}.
We determine an average path, $\boldxgamma$, according to Eq.~\eqref{eq: average path equation} and parameterize its tubular neighborhood with coordinates $s$ and $h$ as in Appendix~\ref{appendix:diffgeo}.
Note that $h$ represents the normal deviation from $\boldxgamma$.
For a trajectory $\boldx(t)$, we use the evolution of its $h$-coordinate, $h(t)$, to determine the distance from $\boldxgamma$:
\begin{equation*}
    \lVert \bar \boldx - \boldx\rVert := \frac{1}{t_2-t_1}\int_{t_1}^{t_2}|h(t)|\,\mathrm{d}t,
\end{equation*}
where the trajectory is defined for time $t\in[t_1,t_2]$.
We improve the bundle by filtering out the 5\% most deviating trajectories. That is $\lVert \bar \boldx - \boldx\rVert>24.6\,\mathrm{cm}$, as depicted in Fig.~\ref{fig: trajectory selection Amsterdam}-\ref{fig:data selection AMS deviation}.

\noindent
\textbf{Trajectory selection EHV.} In contrast with the measurements at Amsterdam train station, nearly always more than one pedestrian is measured at the measurement domain in Eindhoven train station. Therefore we employ a rectangular grid consisting of $3\,\mathrm{m}\times 3\,\mathrm{m}$ cells. We define the local density as the number of pedestrians in a grid cell. To ensure diluted conditions, we only consider trajectories where the local density does not exceed one during their course. Furthermore, we ensure walking trajectories by applying the same velocity restriction as in AMS trajectory selection. 

We group trajectories that originate and terminate in the same areas of the train station into bundles. Five bundles are suited for our analysis as they contain many (curved) paths.
Average paths are determined as before. In a similar fashion to AMS trajectory selection, we improve each bundle by discarding the most deviating 10\%. The average paths of three bundles are displayed in Fig.~\ref{fig:trajectories trainstation} (paths in red, blue, green correspond to bundle 1, 2 and 3 respectively).  
The average paths of bundle 4 and 5 are displayed in Fig.~\ref{fig:remaining sets ehv}.

\section{Construction of tubular neighborhood and derivation of the covariant derivative}\label{appendix:diffgeo}

\subsection*{Covariant derivative}

A covariant derivative (a.k.a. affine connection) is a mapping that describes how vectors change when transporting them in a smooth collection of tangent spaces. The concept of covariant derivative can be understood as an generalization of the ordinary derivative towards curved surfaces.
For $\mathbf{u}$ and $\mathbf{v}$ vectors in a tangent space of a curved surface, the covariant derivative of $\mathbf{u}$ along $\mathbf{v}$ is denoted as $\nabla_{\mathbf{v}}\mathbf{u}$ and respects the following properties (e.g.~\cite{jost2008}): 
\begin{itemize}
    \item[(i)] $\nabla_{f_1\mathbf{v}_1+f_2\mathbf{v}_2}\mathbf{u} = f_1\nabla_{\mathbf{v}_1}\mathbf{u} + f_2\nabla_{\mathbf{v}_2}\mathbf{u}$,
    \item[(ii)] $\nabla_\mathbf{v}(\mathbf{u}_1 + \mathbf{u}_2) = \nabla_\mathbf{v}\mathbf{u}_1+\nabla_\mathbf{v}\mathbf{u}_2$,
    \item[(iii)] $\nabla_\mathbf{v}(f\mathbf{u}) = f\nabla_\mathbf{v}\mathbf{u} + \mathbf{v}(f)\cdot\mathbf{u}$,
\end{itemize}
for $\mathbf{u},\mathbf{u}_1,\mathbf{u}_2, \mathbf{v},\mathbf{v}_1,\mathbf{v}_2$ in a tangent space and $f,f_1,f_2$ smooth functions.

We can define the covariant derivative by defining \textit{Christoffel symbols of the second kind}, $\Gamma_{i j}^k$.
These coefficients determine how basis vectors in different spaces are connected via
\begin{equation}
    \Gamma_{i j}^{k} \mathbf{e}_{k} := \nabla_{\mathbf{e}_{j}} \mathbf{e}_{i}.
\end{equation}
Note that from now on, we will use the Einstein summation convention (e.g.~$\Gamma_{i j}^k\eunit_k\equiv\sum_k \Gamma_{i j}^k\eunit_k$). Using the properties above, we could write the covariant derivative in terms of Christoffel symbols: 
\begin{equation}\label{eq: covariant derivative}
    \nabla_{\mathbf{v}} \mathbf{u}=\frac{\partial \mathbf{u}}{\partial \mathbf{v}}+u^{k} u^{j} \Gamma_{k j}^{i} \mathbf{e}_{i},
\end{equation}
where $\mathbf{u}=u^i\eunit_i$.

The covariant derivative can be pushed forward to other coordinate charts using the coordinate transformation $\phi = \psi_\beta \circ \psi_\alpha^{-1}$, which maps from chart $\psi_\alpha$ to chart $\psi_\beta$. This induces a relation between Christoffel symbols in different coordinate charts:
\begin{equation}\label{eq: push forward christoffel symbols}
    \Gamma_{i j}^{k}=T_{\ell}^{k}\left(S_{j}^{m} S_{i}^{n} \bar{\Gamma}_{n m}^{\ell}+\partial_{j} S_{i}^{\ell}\right),
\end{equation}
with $T=J_{\phi}$ and $S=J_{\phi^{-1}}=\left[J_{\phi}\right]^{-1}$ the (inverse) Jacobian of $\phi$ and $\Gamma_{i j}^{k}$ and $\bar{\Gamma}_{i j}^{k}$ the Christoffel symbols in the coordinate charts $\psi_\alpha$ and $\psi_\beta$ respectively.

\subsection*{Tubular neighborhood}\label{Appendix: tubular neighborhood}
We construct a coordinate chart, $\psi_{\boldxbar}$, that covers the tubular neighborhood of a generic curve $\boldxbar:\mathbb{R}\rightarrow\mathbb{R}^2:s\mapsto\left(x,y\right)$ by using the tangent and normal vectors,
\begin{equation}\label{eq: tangent and normal vector}
    \eunit_\|=\frac{\boldxbar^{\prime}(s)}{\left|\boldxbar^{\prime}(s)\right|}\hspace{1pc} \mathrm{and} \hspace{1pc}
    \eunit_\perp= \left(\begin{array}{cc}
        0 & 1\\ -1 & 0
\end{array}\right) \eunit_\|,
\end{equation}
as basis vectors. The coordinate lines are parallel and normal to $\boldxgamma$ with coordinates $s$ and $h$ representing the parallel and transversal direction respectively. The coordinate transformation form $\psi_{\boldxbar}$ to the Cartesian coordinates is given by
\begin{equation}\label{eq: coordinate transformation}
    \boldsymbol \phi(s, h)=\boldxgamma+h\,\eunit_\perp(s).
\end{equation}

\subsection*{Energy conserving connection}
Geodesics are generally defined as parallel transport of velocity vectors in their own direction \cite{jost2008},
\begin{equation}
    \nablauu=\Ddot{\boldsymbol x}+\Gamma^i_{k j}\Dot{x}^k\Dot{x}^j\mathbf{e}_i=\boldsymbol 0,
\end{equation}
analogously to Eq.~\eqref{eq:geodesic flow} with correction term $C\left(\boldx, \dot{\boldx}, \boldxgamma \right)=\Gamma^i_{k j}\Dot{x}^k\Dot{x}^j\mathbf{e}_i$.
We derive our affine connection (i.e.~derive the Christoffel symbols) such that geodesics respect the physical properties:
\begin{itemize}
    \item geodesic curves conserve kinetic energy;
    \item geodesic curves initially parallel to $\boldxgamma$ remain parallel to $\boldxgamma$ at all times,
\end{itemize}
as explained in Sect.~\ref{section: Langevin-like model for curved tubular neighborhood}.
These properties fully describe geodesics in flat space nearby straight paths as $\ddot{\boldx}=0$ ($\Gamma^i_{k j}=0~\forall_{i,j,k}$). However, this simple connection does not hold for curved paths or curvilinear coordinates.

Energy conservation is ensured by conserving the physical velocity,
\begin{equation}\label{eq:physical velocity}
\lVert \boldsymbol v \rVert^2 = \Tilde{g}_{ij}\dot{q^i}\dot{q^j}, 
\end{equation}
where trajectory $\left(q^1(t), q^2(t)\right)$ is in generic coordinate chart $\psi_q$ with metric $\Tilde{g}$.
By defining the metric tensor in the Cartesian coordinate chart as $g_{ij}=\delta_{ij}$, we define the physical velocity to be the Euclidean velocity ($||\boldsymbol v||^2=\dot{x}^2+\dot{y}^2$). We define coordinate chart $\psi_{\boldxbar}$ as in Appendix~\ref{Appendix: tubular neighborhood}. Then the metric in $\psi_{\boldxbar}$ is given by \cite{jost2008}:
\begin{equation}
\hat{g}_{k q}=g_{i j} \frac{\partial \phi^{i}}{\partial s^{k}} \frac{\partial \phi^{j}}{\partial s^{q}},
\end{equation}
where $\phi$ denotes the coordinate transformation to the Cartesian coordinates (Eq.~\eqref{eq: coordinate transformation}) and $(s^1,s^2)=(s,h)$.
Note that $\hat{g}_{sh}=\hat{g}_{hs}=\langle\eunit_\|,\eunit_\perp\rangle=0$ since $\eunit_\| \perp \eunit_\perp$. Furthermore, $\hat{g}_{hh}=\lVert\eunit_\perp\rVert^2=1$ by definition. Therefore the metric in coordinate chart $\psi_{\boldxbar}$ can be written as
\begin{equation}\label{eq:metric gamma chart}
    \hat{g}_{ij} = \left(\begin{array}{cc}
\left(\partial_s \phi_x\right)^{2}+\left(\partial_s \phi_y\right)^{2} & 0 \\
0 & 1
\end{array}\right).
\end{equation}
Because the metric is diagonal, the physical velocity can be separated into two orthogonal parts,
\begin{equation}
    v^\| = \sqrt{\hat{g}_{ss}}\dot{s}\quad \text{and}\quad v^\perp = \sqrt{\hat{g}_{hh}}\dot{h},
\end{equation}
which are the longitudinal and transversal velocity components respectively. To meet the properties, both velocity components need to be conserved, meaning
\begin{equation}
    \left\{\begin{array}{l}
    \frac{\mathrm{d}}{\mathrm{d} t}v^\|=\frac{\mathrm{d}}{\mathrm{d} t}\left(\sqrt{\hat{g}_{ss}}\dot{s}\right)=0\\[4pt]
    \frac{\mathrm{d}}{\mathrm{d} t}v^\perp=\ddot{h}=0
    \end{array}\right.,
\end{equation}
which can be elaborated to
\begin{equation}\label{eq: conservation of velo}
\left\{\begin{array}{l}
\ddot{s} + \frac{\partial_s \hat{g}_{ss}}{2\hat{g}_{ss}} \dot{s}^{2}+\frac{\partial_h \hat{g}_{ss}}{2\hat{g}_{ss}} \dot{h} \dot{s}=0 \\
\ddot{h}=0
\end{array}\right.
\end{equation}
Using Eq.~\eqref{eq: covariant derivative}, the Christoffel symbols in $\psi_{\boldxbar}$ can be determined such that Eq.~\eqref{eq: conservation of velo} is respected:
\begin{equation}\label{eq: Christoffel symbols s}
\bar{\Gamma}_{i j}^{s}=\left[\begin{array}{cc}
\frac{\partial_s \hat{g}_{ss}}{2\hat{g}_{ss}} & \frac{\partial_h \hat{g}_{ss}}{2\hat{g}_{ss}} \\[6pt]
\frac{\partial_h \hat{g}_{ss}}{2\hat{g}_{ss}} & 0
\end{array}\right],
\end{equation}
and
\begin{equation}\label{eq: Christoffel symbols h}
\bar{\Gamma}_{i j}^{h}=0.
\end{equation}
Note that we can obtain the Christoffel symbols in the Cartesian coordinate chart by applying Eq.~\eqref{eq: push forward christoffel symbols}.

\section{Numerical simulations}\label{Appendix: numerical simulations}

We integrate Eq~\eqref{eq:main} by using the Runge-Kutta SRI2 algorithm~\cite{Rossler2010} (via the PyPI library \texttt{sdeint}~\cite{aburn2016sdeint}). We choose a discretization step size of $0.1\,\mathrm{s}$, similar to the sampling frequency of our measurements.
We initialize our simulations at the beginning of our preferred path $\boldxbar(s)$ with $s\left(t=0\right)=0$ and $h(0)$, $v^\perp(0)$ and $\vshift(0)$ distributed according to the Fokker-Planck equilibrium distributions (see Appendix~\ref{Appendix: model calibration}). 

\begin{figure}[H]
\centering
\begin{tikzpicture}[scale=0.6]
    \draw[fill=green!20] (0,0) rectangle (3.5,6);
    \draw[fill=blue!20] (4,0) rectangle (7.5,6);
    \draw[fill=blue!10] (5,5.6) rectangle (6.5,6.4);
    \draw[fill=green!10] (1,5.6) rectangle (2.5,6.4);

    \node (xy) at (1.75, 6) {$\left(x,\,y\right)$};
    \node (sh) at (5.75, 6) {$\left(s,\,h\right)$};

    \node (CC) at (1.75,5) {$x,\,y$};
    \node (TC) at (5.75,5) {$s,\,h$};

    \node (SS) at (5.75,3.5) {$\mathbf{e}_\|,\,\mathbf{e}_\perp$};
    \node (SS2) at (5.75,2.8) {$\hat{g}_{ij}$};
    \node (CST) at (5.75,1) {\Large $\overline{\Gamma}^i_{jk}$};
    \node (CSC) at (1.75,1) {\Large ${\Gamma}^i_{jk}$};

    \draw[->, thick] (CC.east) -- (TC.west);
    \draw[->, thick] (TC.south) -- (SS.north);
    \draw[->, thick] (SS2.south) -- (CST.north);
    \draw[->, thick] (CST.west) -- (CSC.east);
\end{tikzpicture}
    \caption{A schematic overview of the calculation steps for determining the Christoffel symbols. \textbf{(1)}: Estimation of tubular coordinates using the Newton-Raphson method. \textbf{(2)}: Calculation of tubular neighborhood-dependent variables such as the longitudinal and transversal directions, $\eunit_\|$ and $\eunit_\perp$, and the metric $\hat{g}_{ij}$. \textbf{(3)}: Computation of the Christoffel symbols in the coordinate chart $\psi_{\boldxbar}$. \textbf{(4)}: Push-forward of the Christoffel symbols to the Cartesian coordinates.}\label{fig:simulation scheme}
\end{figure}
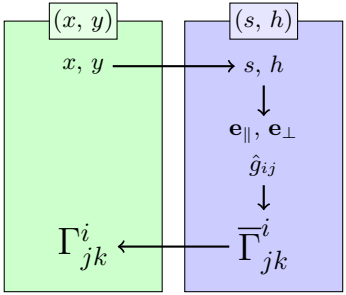

The Christoffel symbols, needed every time step during the integration of Eq.~\eqref{eq:main}, are obtained by the computational steps shown in Fig.~\ref{fig:simulation scheme}. For step~\textbf{(1)}, the computation of the tubular coordinates, we use the two-dimensional Newton-Raphson method~\cite{BenIsrael1966}. This iterative method solves equations of the form $\mathbf{f}(\boldsymbol{s})=0$. If $\boldsymbol{s}_0$ is an approximate solution, then the sequence 
\begin{equation*}
\boldsymbol{s}_{p+1}=\boldsymbol{s}_p-J^{-1}(\boldsymbol{s}_p)\mathbf{f}(\boldsymbol{s}_p)
\end{equation*}
for $p=1,2,...$ and $J$ Jacobian of $\mathbf{f}$, converges to a solution. Given $\boldx$, the tubular coordinates are represented by the roots of function $\mathbf{f}(\mathbf{s})=\phi(\mathbf{s})-\hat{x}$. The roots of $\mathbf{f}$ are estimated with the Newton-Raphson method with the tubular coordinates of the previous time step as an approximated solution. With our typical simulation duration and discretization step size, two iterations of the Newton-Raphson method give a sufficient accurate estimation of coordinates $s$ and $h$.

In step~\textbf{(2)}, we use Eq.~\eqref{eq: tangent and normal vector}, Eq.~\eqref{eq: coordinate transformation} and Eq.~\eqref{eq:metric gamma chart} to calculate the metric in tubular coordinates, $\hat{g}_{ss}$, and the derivatives with respect to $s$ and $h$ ($\partial_s \hat{g}_{ss}$ and $\partial_h \hat{g}_{ss}$). We compute the Christoffel symbols in the tubular coordinate chart in step~\textbf{(3)} using Eq.~\eqref{eq: Christoffel symbols s}-\eqref{eq: Christoffel symbols h}. Finally, in step~\textbf{(4)}, we push the Christoffel symbols to the Cartesian coordinate chart using Eq.~\eqref{eq: push forward christoffel symbols}.

\begin{figure*}[ht]
    \centering
    \begin{subfigure}{0.325\textwidth}
    \includegraphics[width=\textwidth]{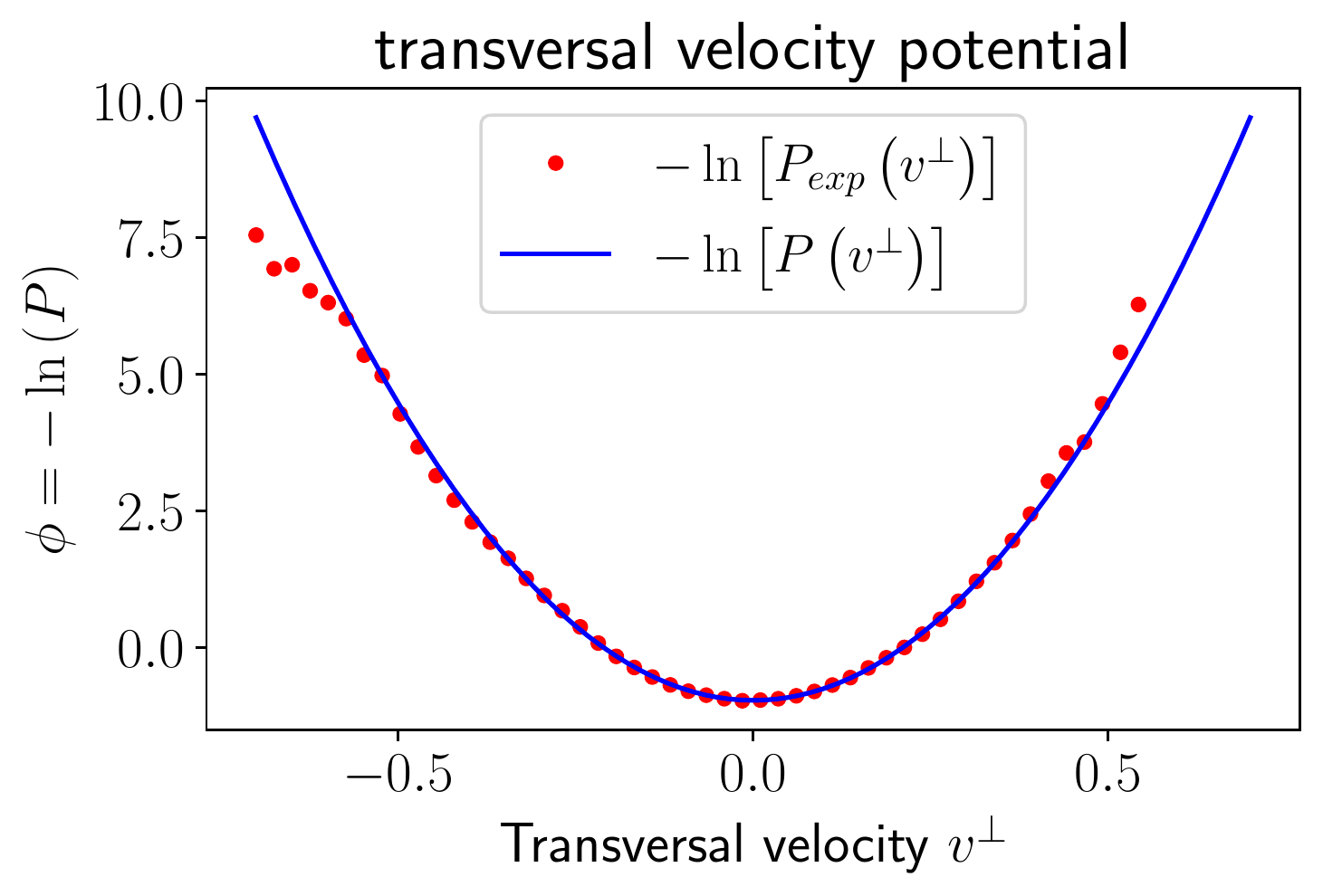}
    \caption{}
    \label{fig: Vperp potential}
    \end{subfigure}
    \hfill
    \begin{subfigure}{0.325\textwidth}
    \includegraphics[width=\textwidth]{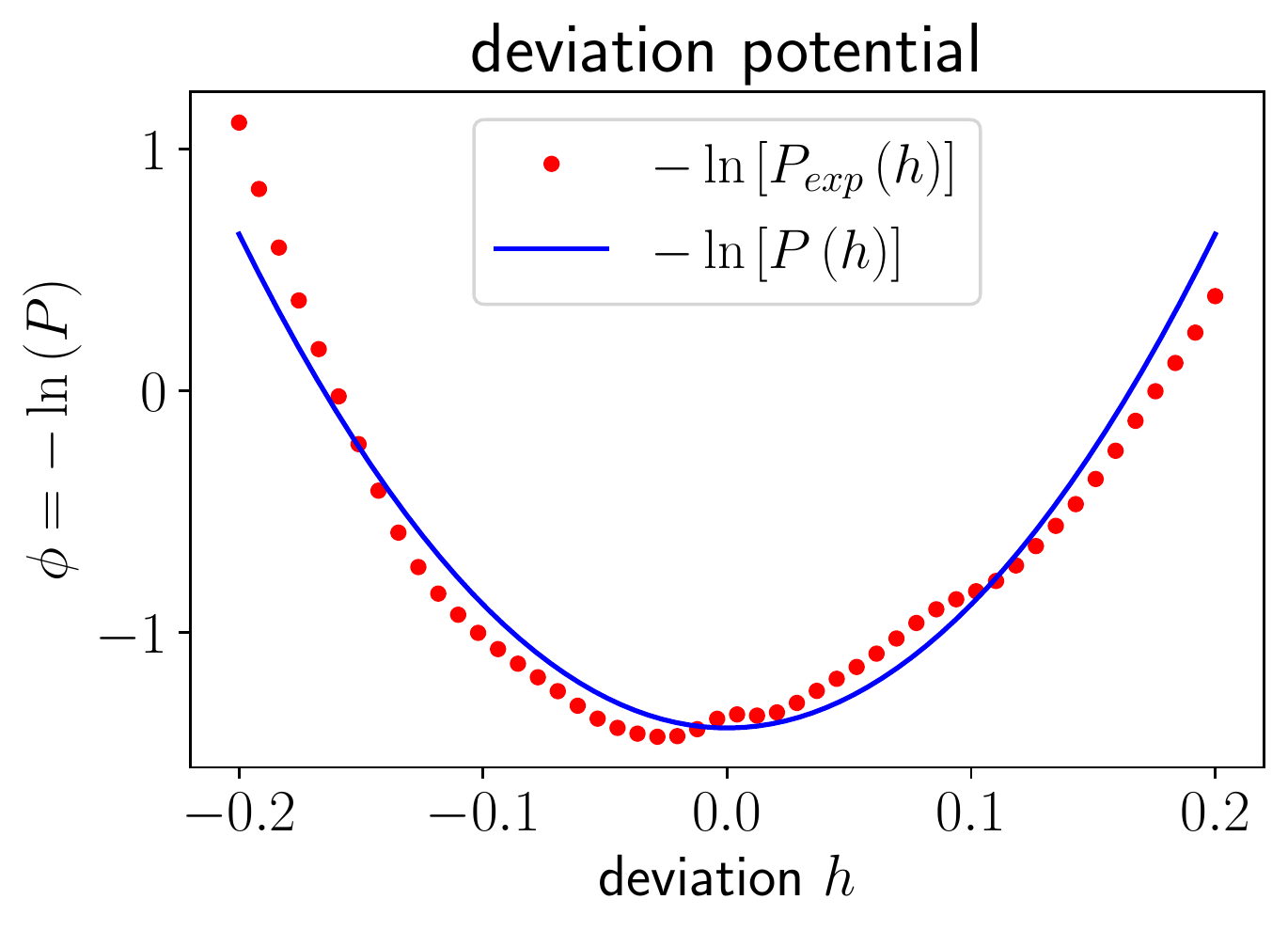}
    \caption{}    
    \label{fig: h potential}
    \end{subfigure}
    \hfill
    \begin{subfigure}{0.325\textwidth}
    \includegraphics[width=\textwidth]{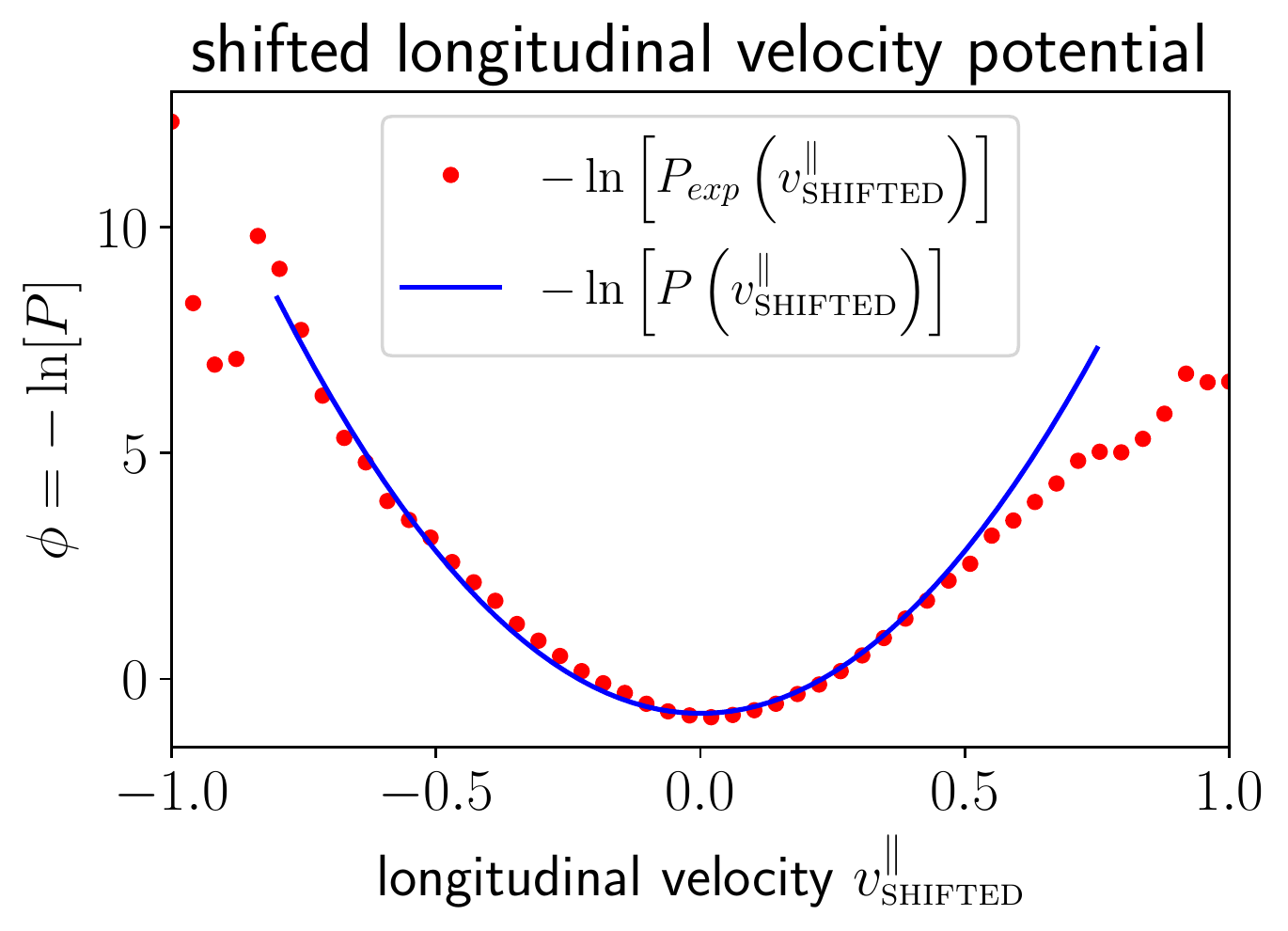}
    \caption{}
    \label{fig: Vparallel shifted potential}
    \end{subfigure}
    \caption{(a) The empiric potentials, $\Phi_{exp}=-\ln{\left(\Prob_{exp}\right)}$, of (a) the transversal velocity, (b) the lateral deviation and (c) the shifted longitudinal velocity obtained from the Amsterdam train station data (red) compared with the fitted model potentials (blue).}
    \label{fig: potentials with fits}
\end{figure*}
\section{Model calibration}\label{Appendix: model calibration}
The model is calibrated by estimating the model parameters, $\{\alpha, \beta, \mu, \vSP, \delta, \sigma\}$. We use the Amsterdam train station measurements to estimate the parameters. 
The ‘straight-path velocity’ $\vSP$ is estimated by linearly extrapolating the $v^\| - k$ relation towards $k=0$.  For the Amsterdam train station this results in $\vSP=1.33\,\mathrm{ms}^{-1}$. The body size radius, $\delta$, represent the slope of the fundamental diagram (Fig.~\ref{fig: fundamental diagram}). The estimation of $\delta$ is obtained by a linear fit: $\delta=0.19\,\mathrm{m}$.
The remaining four parameters are estimated by applying fits to empirical Langevin potentials and a correlation function. The first fit is applied to the transversal velocity potential. In the stationary regime, the model produces probability distribution of the transversal velocity and lateral deviation from the preferred path, $\Prob (h,v^\perp)$, according the well-known Fokker-Planck equation \cite{Palacios2007} with solutions
\begin{equation}
\Prob\left(h, v^\perp\right)=\mathcal{N} \exp \left[-\frac{2 \mu}{\sigma^{2}} \left({v^\perp}\right)^{2}-\frac{4 \beta \mu}{\sigma^{2}} h^{2}\right]
\end{equation}
where $\mathcal{N}$ denotes a normalization constant. A Langevin potential can be constructed according to $\Phi(\cdot)=-\ln{\left(\Prob(\cdot)\right)}$. The analytical potentials of the transversal dynamics should agree with the empiric potentials such that
\begin{equation}
-\ln{\left(\Prob_{exp}(v^\perp)\right)} \approx \frac{2\mu}{\sigma^2}\left({v^\perp}\right)^2 + K_1
\end{equation} 
and
\begin{equation}
-\ln{\left(\Prob_{exp}(h)\right)} \approx \frac{4\beta\mu}{\sigma^2}h^2 + K_2.
\end{equation}
The constants $K_1$ and $K_2$ are normalization constants and $\Prob_{exp}(\cdot)$ denotes the empiric probability distribution function. The fitting can be observed in Fig.~\ref{fig: Vperp potential}-\ref{fig: h potential} where the resulting estimated ratios are given by
\begin{equation}\label{eq: fit transversal potentials}
\frac{2\mu}{\sigma^2}\approx 21.77 \quad \mathrm{and} \quad \frac{4\beta\mu}{\sigma^2}\approx 51.08.
\end{equation}

The same can be done for the longitudinal dynamics. In the stationary regime, the probability of the shifted longitudinal velocity is distributed according to
\begin{equation}
\Prob\left( \vshift \right)=\mathcal{N} \exp \left[\frac{2 \alpha }{\sigma^{2}} \left({\vshift}\right)^{2}\right]\end{equation}
where $\mathcal{N}$ is a normalization constant. The ratio $\frac{2\alpha}{\sigma^2}$ is compared to the empirical distribution function of the shifted longitudinal velocity according to
\begin{equation}
-\ln{\left(\Prob_{exp}( \vshift )\right)} \approx \frac{2\alpha}{\sigma^2} \left(\vshift\right)^2 +K_3.
\end{equation}
Constant $K_3$ again represents normalization. The fit (Fig.~\ref{fig: Vparallel shifted potential}) results in the estimation of the ratio:
\begin{equation}\label{eq: fit V longitudinal shifted pot}
\frac{2\alpha}{\sigma^2}\approx 14.36.
\end{equation}
To complete the parameter estimation, a time correlation function of the shifted longitudinal velocity is used. Using Eq.~\eqref{eq:fvvbc} and the definition of $\vshift$, the deterministic shifted longitudinal dynamics can be described by
\begin{equation}
    \frac{\mathrm{d}}{\mathrm{d}t}\vshift=-2\alpha \vshift.
\end{equation}
Therefore, the time correlation of $\vshift$ should decay as $\exp{\left(-2\alpha t\right)}$. An estimated value of $\alpha$ follows from the fit (Fig.~\ref{fig: V shifted correlation AMS}):
\begin{equation}\label{eq: fit time cor}
    -2\alpha \approx -0.51
\end{equation}

The estimates obtained by the fitted values result in the parameter values reported in \autoref{table: errors parameter estimation}. 
To determine uncertainty intervals for our estimates, we repeat the fitting procedure five times using randomly selected, equally-sized partitions of the data. 
We then use the fitted values from each of the five partitions to estimate the minimum and maximum values for each parameter. We set these as the lower and upper bounds of the respective intervals.

\begin{table}[H]

\caption{Estimated parameter values with associated uncertainty intervals.}
\begin{center}
\begin{tabular}{l|l|c|l}
\hline
parameter             & value & uncertainty interval & \\ \hline
$\alpha$              & 0.26  & $[0.22,\, 0.28]$ & s$^{-1}$\\
$\beta$               & 1.17   & $[0.80,\, 1.67]$ & s$^{-2}$\\
$\mu$                 & 0.39  & $[0.31,\, 0.46]$ &s$^{-1}$\\
$\sigma$              & 0.19  & $[0.17,\, 0.20]$ & ms$^{-3/2}$\\
$v_\textsc{\tiny{SP}}$       & 1.33 & $[1.29,\, 1.35]$ & ms$^{-1}$\\
$\delta$              & 0.192 & $[0.187,\, 0.195]$ & m\\ \hline
\end{tabular}\label{table: errors parameter estimation}
\end{center}
\end{table}

\bibliographystyle{apsrev4-1}

\bibliography{references}
\end{document}